\documentclass[12pt, authoryear]{article}
\usepackage{amsfonts, graphicx,color,multirow, amsthm,amsmath,natbib, hypernat}
\usepackage{setspace,url}
\usepackage{multirow}

\def\boxit#1{\vbox{\hrule\hbox{\vrule\kern6pt\vbox{\kern6pt#1\kern6pt}\kern6pt\vrule}\hrule}}

\begin{document}
\newtheorem{theorem}{Theorem}
\newtheorem{corollary}{Corollary}
\newtheorem{case}{Case}
\newtheorem{definition}{Definition}
\newtheorem{example}{Example}
\newtheorem{lemma}{Lemma}
\newtheorem{proposition}{Proposition}
\newtheorem{remark}{Remark}
\newtheorem{assumption}{Assumption}

\newcommand{\ol}[1]{\overline{#1}}

\newcommand{\bb}{\mbox{\bf b}}
\newcommand{\bff}{\mbox{\bf f}}
\newcommand{\bx}{\mbox{\bf x}}
\newcommand{\by}{\mbox{\bf y}}
\newcommand{\bA}{\mbox{\bf A}}
\newcommand{\ba}{\mbox{\bf a}}
\newcommand{\bw}{\mbox{\bf w}}
\newcommand{\bu}{\mbox{\bf u}}
\newcommand{\bB}{\mbox{\bf B}}
\newcommand{\bC}{\mbox{\bf C}}
\newcommand{\bD}{\mbox{\bf D}}
\newcommand{\bE}{\mbox{\bf E}}
\newcommand{\bF}{\mbox{\bf F}}
\newcommand{\bG}{\mbox{\bf G}}
\newcommand{\bH}{\mbox{\bf H}}
\newcommand{\bI}{\mbox{\bf I}}
\newcommand{\bK}{\mbox{\bf K}}
\newcommand{\bL}{\mbox{\bf L}}
\newcommand{\bQ}{\mbox{\bf Q}}
\newcommand{\bR}{\mbox{\bf R}}
\newcommand{\bS}{\mbox{\bf S}}
\newcommand{\bU}{\mbox{\bf U}}
\newcommand{\bX}{\mbox{\bf X}}
\newcommand{\bW}{\mbox{\bf W}}
\newcommand{\bY}{\mbox{\bf Y}}
\newcommand{\bZ}{\mbox{\bf Z}}
\newcommand{\bone}{\mbox{\bf 1}}
\newcommand{\bsone}{\mbox{\bf 1}}
\newcommand{\bzero}{\mbox{\bf 0}}
\newcommand{\bveps}{\mbox{\boldmath $\varepsilon$}}
\newcommand{\bet}{\mbox{\boldmath $\eta$}}
\newcommand{\bxi}{\mbox{\boldmath $b$}}
\newcommand{\beps}{\mbox{\boldmath $\varepsilon$}}
\newcommand{\bmu}{\mbox{\boldmath $\mu$}}
\newcommand{\bgamma}{\mbox{\boldmath $\gamma$}}
\newcommand{\bbeta}{\mbox{\boldmath $\beta$}}
\newcommand{\bGamma}{\mbox{\boldmath $\Gamma$}}
\newcommand{\mv}{\mbox{V}}
\newcommand{\bSigma}{\mbox{\boldmath $\Sigma$}}
\newcommand{\bOmega}{\mbox{\boldmath $\Omega$}}
\newcommand{\FDR}{\mbox{FDR}}
\newcommand{\FDP}{\mbox{FDP}}
\newcommand{\wFDP}{\mbox{$\widehat{FDP}$}}
\newcommand{\wFDR}{\mbox{$\widehat{FDR}$}}
\newcommand{\LS}{\mbox{\scriptsize LS}}
\newcommand{\Ll}{\mbox{\scriptsize $L_1$}}
\newcommand{\Var}{\mbox{Var}}
\newcommand{\Cov}{\mbox{Cov}}
\newcommand{\SNP}{\mbox{SNP}}
\newcommand{\SD}{\mbox{SD}}

\newcommand{\hB}{\widehat \bB}
\newcommand{\hb}{\widehat \bb}
\newcommand{\hw}{\widehat \bw}
\newcommand{\hE}{\widehat \bE}
\newcommand{\hvar}{\widehat \var}
\newcommand{\hcov}{\widehat \cov}
\newcommand{\hbveps}{\widehat\bveps}
\newcommand{\hSig}{\widehat\Sig}
\newcommand{\hsig}{\widehat\sigma}
\newcommand{\hmu}{\widehat\bmu}
\newcommand{\hxi}{\widehat\bxi}
\newcommand{\heq}{\ \widehat=\ }
\newcommand{\sam}{_{\text{sam}}}
\newcommand{\cov}{\mathrm{cov}}
\newcommand{\Sig}{\mathbf{\Sigma}}
\newcommand{\veps}{\varepsilon}
\newcommand{\tr}{\mathrm{tr}}
\newcommand{\diag}{\mathrm{diag}}
\newcommand{\argmin}{\mathrm{argmin}}
\newcommand{\vecc}{\mathrm{vec}}
\newcommand{\var}{\mathrm{var}}
\def\cov{\mbox{cov}}
\newcommand{\trace}{\mathrm{tr}}

\def\t{^T}
\def\toD{\overset{\mathrm{D}}{\longrightarrow}}
\def\toP{\overset{\mathrm{P}}{\longrightarrow}}
\def\toas{\overset{\mathrm{a.s.}}{\longrightarrow}}
\def\deq{\overset{\mathrm{(d)}}{=\hspace{-0.02 in}=}}
%%%%%%%%===============End Self-define command=======

\def\blackbox{\vrule height4pt width3pt depth0pt}
\def\indep{\perp \!\!\!\!\perp}
\def\bB{\mbox{\bf B}}

%\begin{frontmatter}
\title{\bf Control of the False Discovery Rate Under Arbitrary Covariance Dependence\thanks{Address Information: Jianqing Fan, Department of Operations Research \& Financial Engineering, Princeton University, Sherrerd Hall, Princeton, NJ 08544, USA. Email: jqfan@princeton.edu. This research was partly supported by NSF Grants DMS-0704337 and DMS-0714554 and NIH Grant R01-GM072611.}
}

\author{Xu Han, Weijie Gu and Jianqing Fan}
%\date{This version: Dec 8}
\maketitle

\begin{abstract}
\begin{singlespace}
Multiple hypothesis testing is a fundamental problem in high dimensional inference, with wide applications in many scientific fields. In genome-wide association studies, tens of thousands of tests are performed simultaneously to find if any genes are associated with some traits and those tests are correlated. When test statistics are correlated, false discovery control becomes very challenging under arbitrary dependence. In the current paper, we propose a new methodology based on principal factor approximation, which successfully substracts the common dependence and weakens significantly the correlation structure, to deal with an arbitrary dependence structure. We derive the theoretical distribution for false discovery proportion (FDP) in large scale multiple testing when a common threshold is used and provide a consistent FDP. This result has important applications in controlling FDR and FDP. Our estimate of FDP compares favorably with Efron (2007)'s approach, as demonstrated by in the simulated examples. Our approach is further illustrated by some real data applications.
\end{singlespace}
\end{abstract}

\section{Introduction}
Multiple hypothesis testing is a fundamental problem in the modern research for high dimensional inference, with wide applications in scientific fields, such as biology, medicine, genetics, neuroscience, economics and finance. For example, in genome-wide association studies, massive amount of genomic data (e.g. SNPs, eQTLs) are collected and tens of thousands of hypotheses are tested simultaneously to find if any of these genes are associated with some observable traits (e.g. blood pressure, weight, some disease); in finance, thousands of tests are performed to see which fund managers have winning ability (Barras, Scaillet \& Wermers 2010)

False Discovery Rate (FDR) has been introduced in the celebrated paper by Benjamini \& Hochberg (1995) for large scale multiple testing. By definition, FDR is the expected proportion of falsely rejected null hypotheses among all of the rejected null hypotheses. The classification of tested hypotheses can be summarized in Table 1:
\begin{table}[h!!]
\begin{center}\caption{Classification of tested hypotheses}\label{ar}
  \begin{tabular}{cccc}
         \hline\hline
                       &Number    &Number    &\\
         Number of     &not rejected  &rejected  &\\
         \hline
         True Null     &$U$         &$V$      &$p_0$\\
         False Null    &$T$         &$S$      &$p_1$\\
         \hline
                       &$p-R$       &$R$      &$p$\\
         \hline
   \end{tabular}
\end{center}
\end{table}

Various testing procedures have been developed for controlling FDR, among which there are two major approaches. One is to compare the ordered $P$-values respectively with a sequence of threshold values (Benjamini \& Hochberg 1995).  Specifically, let
$p_{(1)} \leq p_{(2)} \leq \cdots \leq p_{(p)}$ be the ordered observed $P$-values of $p$ hypotheses. Define $k = \text{max}\Big\{i: p_{(i)} \leq i\alpha/p \Big\}$ and reject $H_{(1)}^0, \cdots, H_{(k)}^0$, where $\alpha$ is a specified control rate. If no such $i$ exists, reject no hypothesis. The other related approach is to fix a threshold value and reject the hypotheses with $P$-values no greater than this threshold (Storey 2002). The equivalence between the two methods has been theoretically studied by Storey, Taylor \& Siegmund (2004) and Ferreira \& Zwinderman (2006). Finding such a common threshold is based on a conservative estimate of FDR. Specifically, let $\mathrm{\widehat{FDR}}(t) = \widehat{p}_0t/(R(t)\vee 1)$, where $R(t) = \#\{P_i: P_i \leq t\}$ is the number of total discoveries with the threshold $t$ and $\widehat{p}_0$ is an estimate of $p_0$. Then solve $t$ such that $\mathrm{\widehat{FDR}}(t)\leq\alpha$ where $\alpha$ is a predetermined control rate, say $15\%$.

Both procedures have been shown to perform well for independent test statistics. However, in practice, test statistics are usually correlated. Although Clarke \& Hall (2009) argued that when the null distribution of test statistics satisfies some conditions, dependence case in the multiple testing is asymptotically the same as independence case, multiple testing under general dependence structures is still a very challenging and important open problem. Efron (2007) noted that correlation must be accounted for in deciding which null hypotheses are significant because the accuracy of false discovery rate techniques will be compromised in high correlation situations. There are several literatures to show that Benjamini-Hochberg procedure or Storey's procedure can control FDR under some special dependence structures, e.g. Positive Regression Dependence on Subsets (Benjamini \& Yekutieli 2001) and weak dependence (Storey, Taylor \& Siegmund 2004). Sarkar (2002) also shows that FDR can be controlled by a generalized stepwise multiple testing procedure under positive regression dependence on subsets. However, even if the procedures are valid under these special dependence structures, they will still suffer from efficiency loss without considering the actual dependence information. In other words, there are universal upper bounds for a given class of covariance matrices.

In the current paper, we will develop a procedure for high dimensional multiple testing which can deal with any arbitrary dependence structure and fully incorporate the covariance information. This is in contrast with Sun \& Cai (2009) who developed a multiple testing procedure under a hidden Markov model and Leek \& Storey (2008) and Friguet, Kloareg \& Causeur (2009) where the factor models are imposed. More specifically, consider the test statistics
\begin{equation*}
(Z_1,\cdots,Z_p)^T\sim N((\mu_1,\cdots,\mu_p)^T,\bSigma),
\end{equation*}
where $\bSigma$ is known and $p$ is large. We would like to simultaneously test $H_{0i}: \mu_i=0$ vs $H_{1i}: \mu_i\neq0$ for $i=1,\cdots,p$. Note that $\bSigma$ can be any non-negative definite matrix. Our procedure is called Principal Factor Approximation (PFA). The basic idea is to first take out the principal factors that derive the strong dependence among observed data $Z_1,\cdots,Z_p$ and to account for such dependence in FDP calculation. This is accomplished by the spectral decomposition of $\bSigma$ and taking out the largest common factors so that the remaining dependence is weak. We then derive the theoretical distribution of false discovery proportion $V/R$ when $p$ is large that accounts for the strong dependence. The realized but unobserved principal factors that derive the strong dependence are then consistently estimated. We will further discuss the application of our result in multiple testing.

The motivation for this problem setup comes from genome-wide association studies. We are especially interested in the high dimensional sparse problem, that is, $p$ is very large, but the number of $\mu_i\neq0$ is very small. In section 2, we will further explain why $\bSigma$ is known in practice. Sections 3 and 4 present the theoretical results and the proposed procedures. In section 5, the performance of our procedures is critically evaluated by various simulation studies. Section 6 is about the real data analysis. All the proofs are relegated to the Appendix.

\section{Motivation of the Study}
In genome-wide association studies, consider $p$ SNP genotype data for $n$ individual samples, and further suppose that a response of interest (i.e. gene expression level or a measure of phenotype such as blood pressure or weight) is recorded for each sample. The SNP data are conventionally stored in an $n\times p$ matrix $\bX=(x_{ij})$, with rows corresponding to individual samples and columns corresponding to individual SNPs . The total number $n$ of samples is in the order of hundreds, and the number $p$ of SNPs is in the order of tens of thousands.

Let $X_j$ and $Y$ denote, respectively, the random variables that correspond to the $j$th SNP coding and the phenotype. The biological question of the association between genotype and phenotype can be restated as a problem in multiple hypothesis testing, {\em i.e.}, the simultaneous tests for each SNP $j$ of the null hypothesis $H_j$ of no association between the SNP $X_j$ and $Y$. Consider the marginal linear regression between $Y$ and $X_j$:
\begin{equation}\label{gwj1}
\min_{a_j,b_j}E(Y-a_j-b_jX_j)^2, \ \ \ j=1,\cdots,p.
\end{equation}
Let $\alpha_j$ and $\beta_j$ be the solution to (1). We wish to simultaneously test the hypotheses
\begin{equation}
H_{0j}:\quad\beta_j=0\quad\text{vs}\quad H_{1j}:\quad\beta_j\neq0, \quad\quad j=1,\cdots,p
\end{equation}
to see which SNPs are correlated with the phenotype.

Recently statisticians have increasing interests in the high dimensional sparse problem: although the number of hypotheses to be tested is large, the number of false nulls ($\beta_j\neq0$) is very small. For example, among the 2000 SNPs there are maybe only 10 SNPs which contribute to the variation in phenotypes or certain gene expression level. Our purpose is to find out these 10 SNPs by multiple testing with some statistical accuracy.

Because of the correlations among $X_1,\cdots,X_p$, based on a random sample of size $n$, the least-squares estimators $\{\widehat{\beta}_j\}_{j=1}^p$ for $\{\beta_j\}_{j=1}^p$ in (1) are also correlated. The following result describes the joint distribution of $\{\widehat{\beta}_j\}_{j=1}^p$. The proof is straightforward.

{\proposition Let $\widehat{\beta}_j$ be the least-squares estimator for $\beta_j$ in (1) based on $n$ data points, $\widehat{\rho}_{kl}$ be the sample correlation between
$X_k$ and $X_l$, and $\widehat{\sigma}_k$ be the sample standard deviation for $X_k$. Assume that the conditional distribution of $Y$ given $X_1,\cdots,X_p$ is $N(\mu(X_1,\cdots,X_p),\sigma^2)$. Then, conditioning on $\{X_{ij}\}$ the joint distribution of $\{\widehat{\beta}_j\}_{j=1}^p$ is
$(\widehat{\beta}_1,\cdots,\widehat{\beta}_p)^T\sim N((\beta_1,\cdots,\beta_p)^T,\bSigma^*)$, where the $(k,l)$th element in $\bSigma^*$ is $\displaystyle\bSigma_{kl}^*=\sigma^2\widehat{\rho}_{kl}/(n\widehat{\sigma}_k\widehat{\sigma}_l)$.}

For ease of notation, let $Z_1,\cdots,Z_p$ be the standardized random variables of $\widehat{\beta}_1,\cdots,\widehat{\beta}_p$, that is,
\begin{equation} \label{b2}
Z_i=\frac{\widehat{\beta}_i}{\SD(\widehat{\beta}_i)}=\frac{\widehat{\beta}_i}{\sigma/(\sqrt{n}\widehat{\sigma}_i)}, \quad\quad i=1,\cdots,p.
\end{equation}
Then, conditioning on $\{X_{ij}\}$,
\begin{equation}\label{c1}
(Z_1,\cdots,Z_p)^T\sim N((\mu_1,\cdots,\mu_p)^T,\bSigma),
\end{equation}
where $\mu_i=\sqrt{n}\beta_i\widehat{\sigma}_i/\sigma$ and covariance matrix $\bSigma$ has the $(k,l)$th element as $\widehat{\rho}_{kl}$. Simultaneously testing (2) based on $(\widehat{\beta}_1,\cdots,\widehat{\beta}_p)^T$ is thus equivalent to testing

\begin{equation}\label{c2}
H_{0j}:\quad\mu_j=0\quad\text{vs}\quad H_{1j}:\quad \mu_j\neq0, \quad\quad j=1,\cdots,p
\end{equation}
based on $(Z_1,\cdots,Z_p)^T$.

In (4), $\bSigma$ is the population covariance matrix of $(Z_1,\cdots,Z_p)^T$, and is known. The covariance matrix $\bSigma$ can have arbitrary dependence structure. Even if the population correlation matrix of the SNP data has certain dependence structure, $\bSigma$ can still be quite different because the sample size $n$ is relatively small.

\section{Estimating False Discovery Proportion}
From now on assume that among all the $p$ null hypotheses, $p_0$ of them are true and $p_1$ hypotheses ($p_1 = p - p_0$) are
false, and $p_1$ is supposed to be very small compared to $p$. For a fixed rejection threshold $t$, we will reject those $P$-values no greater than $t$ and select them as significance. Because of its powerful applicability, this procedure has been widely adopted by many statisticians. See Storey (2002), Efron (2007, 2010), among others. Our goal is to find a common threshold $t$ such that the decision rule has nice statistical properties in multiple testing problem (\ref{c2}) based on the observations (\ref{c1}) under arbitrary dependence structure of $\bSigma$.

\subsection{Approximation of FDP}
Define the following empirical processes:
\begin{eqnarray*}
V(t) & = & \#\{true \ null \ P_i: P_i \leq t\}, \nonumber\\
S(t) & = & \#\{false \ null \ P_i: P_i \leq t\} \quad \text{and} \nonumber\\
R(t) & = & \#\{P_i: P_i \leq t\}, \nonumber
\end{eqnarray*}
where $t\in[0,1]$. $V(t)$, $S(t)$ and $R(t)$ are the number of false discoveries, the number of true discoveries, and the number of total discoveries, respectively. Obviously, $R(t) = V(t) + S(t)$, and $V(t)$, $S(t)$ and $R(t)$ are all random variables, due to the randomness of the test statistics $(Z_1,\cdots,Z_p)^T$. Moreover, $R(t)$ is observed given some threshold value $t$ in an experiment, but $V(t)$ and $S(t)$ are both unobserved.

By definition, $\FDP(t)=V(t)/R(t)$ and $\FDR(t) = E\Big[V(t)/R(t)\Big]$. The goal is to control FDR$(t)$ at a predetermined rate $\alpha$, say $15\%$. There are also substantial research interests in the statistical behavior of the number of false discoveries $V(t)$ and the false discovery proportion $V(t)/R(t)$, which are unknown but realized.

We will explore the distribution of $V(t)/R(t)$ for the high dimensional sparse case $p_1\ll p$. Suppose $(Z_1,\cdots,Z_p)^T\sim N((\mu_1,\cdots,\mu_p)^T,\bSigma)$. The covariance matrix $\bSigma$ has the $(k,l)$th element as $\rho_{kl}$ with $\rho_{kk}=1$ so that it is a correlation matrix. $\bSigma$ can be any non-negative definite matrix. Our setting encompasses the problem in Section 2. Before we introduce our procedure to deal with the arbitrary dependence case, we will give the following definition for weakly dependent normal random variables, which is fundamental to our method.
\begin{definition}
Suppose $(K_1,\cdots,K_p)^T\sim N((\theta_1,\cdots,\theta_p)^T,\bA)$. Then $K_1,\cdots,K_p$ are called weakly dependent normal variables if
\begin{equation}
\lim_{p\rightarrow\infty}p^{-2}\sum_{i,j}|a_{ij}|=0,
\end{equation}
where $a_{ij}$ denote the $(i,j)$th element of covariance matrix $\bA$.
\end{definition}

Our procedure is called principal factor approximation (PFA). The basic idea is that any $(Z_1,\cdots,Z_p)^T\sim N((\mu_1,\cdots,\mu_p)^T,\bSigma)$ can be decomposed as a factor model with weakly dependent normal random errors. The details are shown as follows.
Firstly apply the spectral decomposition to the covariance matrix $\bSigma$. Suppose the eigenvalues of $\bSigma$ are $\lambda_1,\cdots,\lambda_{p}$, which have been arranged in decreasing order. If the corresponding orthonormal eigenvectors are denoted as $\bgamma_1,\cdots,\bgamma_p$, then
\begin{equation}
\bSigma=\sum_{i=1}^p\lambda_i\bgamma_i\bgamma_i^T.
\end{equation}
If we further denote $\bA = \sum_{i=k+1}^p\lambda_i\bgamma_i\bgamma_i^T$ where $k$ is some well-chosen integer value, then
\begin{equation}
\|\bA\|_F^2 =\lambda_{k+1}^2+\cdots+\lambda_p^2,
\end{equation}
where $\|\cdot\|_F$ is the Frobenius norm. Let $\bL=(\sqrt{\lambda_1}\bgamma_1,\cdots,\sqrt{\lambda_k}\bgamma_k)$, which is $p\times k$ dimensional. Then the covariance matrix $\bSigma$ can be expressed as
\begin{equation}
\bSigma=\bL\bL^T+\bA,
\end{equation}
and $Z_1,\cdots,Z_p$ can be written as
\begin{equation} \label{b20}
Z_i=\mu_i+\sum_{h=1}^kb_{ih}W_h+K_i, \quad\quad i=1,\cdots,p,
\end{equation}
where $(b_{1j},\cdots,b_{pj})^T=\sqrt{\lambda_j}\bgamma_j$, the factors are $W_h\sim N(0,1)$ and the random errors are $ (K_1,\cdots,K_p)^T$ $\sim N(0,\bA)$. Furthermore, $W_1,\cdots,W_k$ are independent of each other and independent of $K_1,\cdots,K_p$. In expression (\ref{b20}), $\{\mu_i=0\}$ correspond to the true null hypotheses, while $\{\mu_i\neq0\}$ correspond to the false ones. Note that although (\ref{b20}) is not exactly a classical multifactor model because of the existence of dependence among $K_1,\cdots,K_p$, we can nevertheless show that $(K_1,\cdots,K_p)^T$ is a weakly dependent vector if the number of factors $k$ is appropriately chosen.

We now discuss how to choose $k$ such that $(K_1,\cdots,K_p)^T$ is weakly dependent. Denote by $a_{ij}$ the $(i,j)$th element in the covariance matrix $\bA$. If we have
\begin{equation}\label{c3}
p^{-1}(\lambda_{k+1}^2+\cdots+\lambda_p^2)^{1/2} \longrightarrow 0 \ \text{as} \ p\rightarrow\infty,
\end{equation}
then
\begin{equation*}
p^{-2}\sum_{i,j}|a_{ij}|\leq p^{-1}\|\bA\|_F=p^{-1}(\lambda_{k+1}^2+\cdots+\lambda_p^2)^{1/2}\longrightarrow0 \ \text{as} \ p\rightarrow\infty,
\end{equation*}
where the first inequality is by the Cauchy-Schwartz inequality. Note that $\sum_{i=1}^p\lambda_i=tr(\bSigma)=p$, so that (\ref{c3}) is self-normalized. Therefore, by definition $(K_1,\cdots,K_p)^T$ is weakly dependent. In practice, we always choose the smallest $k$ such that
\begin{equation*}
\frac{\sqrt{\lambda_{k+1}^2+\cdots+\lambda_{p}^2}}{\lambda_1+\cdots+\lambda_p} < \varepsilon
\end{equation*}
holds for a predetermined small $\varepsilon$, say, $0.01$.

\begin{theorem}
Suppose $(Z_1,\cdots,Z_p)^T\sim N((\mu_1,\cdots,\mu_p)^T,\bSigma)$. Choose an appropriate $k$ such that
\begin{equation*}
(C0) \ \ \ \ \ \ \ \ \frac{\sqrt{\lambda_{k+1}^2+\cdots+\lambda_{p}^2}}{\lambda_1+\cdots+\lambda_p}=O(p^{-\delta}) \ \ \ \text{for} \ \ \delta>0.
\end{equation*}
Let $\sqrt{\lambda_j}\bgamma_j=(b_{1j},\cdots,b_{pj})^T$ for $j=1,\cdots,k$ and $(W_1,\cdots,W_k)^T\sim N_k(0,\bI_k)$. Then,
\textrm{
\begin{equation}\label{a50}
\lim_{p_0\rightarrow\infty}\mathrm{FDP}(t)\stackrel{D}{=}\frac{\sum_{i\in\text{\{true null\}}}\Big[\Phi(a_i(z_{t/2}+\eta_i))+\Phi(a_i(z_{t/2}-\eta_i))\Big]}{\sum_{i=1}^p\Big[\Phi(a_i(z_{t/2}+\eta_i+\mu_i))+\Phi(a_i(z_{t/2}-\eta_i-\mu_i))\Big]},
\end{equation} }
where $a_i = (1-\sum_{h=1}^kb_{ih}^2)^{-1/2}$, $\eta_i = \sum_{h=1}^kb_{ih}W_h$, and $\Phi(\cdot)$ and $z_{t/2}=\Phi^{-1}(t/2)$ are the cumulative distribution function and the $t/2$ lower quantile of a standard normal distribution, respectively.
\end{theorem}
Note that condition (C0) implies that $K_1,\cdots,K_p$ are weakly dependent random variables, but (\ref{c3}) converges to zero at some polynomial rate of $p$.

The result of the asymptotic distribution of FDP$(t)$ in Theorem 1 is new, compared with the current research in multiple testing for general dependence structure. To the best of our knowledge, it is the first result to fully capture the behavior of FDP$(t)$ for high dimensional sparse problem, and the impact of dependence is explicitly spelled out. It is also closely connected with the existing results for independence case and weak dependence case. Let $b_{ih}=0$ for $i=1,\cdots,p$ and $h=1,\cdots,k$ in (\ref{b20}) and $K_1,\cdots,K_p$ are weakly dependent or independent normal random variables, then it reduces to the weak dependence case or independence case, respectively. In the above two special cases, the numerator of (\ref{a50}) is just $p_0t$. Storey (2002) used an estimate for $p_0$, resulting an estimator of $\FDP(t)$ as $\widehat{p}_0t/R(t)$. This estimator has been shown to perform well under independency and weak dependency. However, for general dependency, Storey's procedure will not work well because it ignores the correlation effect among the test statistics, as shown by (\ref{a50}). Further discussions for the relationship between our result and the other leading research for multiple testing under dependence are shown in Section 3.4.

The results in Theorem 1 can be better understood by some special dependence structures as follows. These specific cases are also considered by Roquain \& Villers (2010) and Friguet, Kloareg \& Causeur (2009) under somewhat different setting, but results in Examples 1 and 2 are new.

\textbf{Example 1: [Equal Correlation]}  If $\bSigma$ has $\rho_{ij}=\rho\in[0,1)$ for $i\neq j$, then we can write
\begin{equation*}
Z_i=\mu_i+\sqrt{\rho}W+\sqrt{1-\rho}K_i \ \ \ i=1,\cdots,p
\end{equation*}
where $W\sim N(0,1)$, $K_i\sim N(0,1)$, and $W$ and all $K_i$'s are independent of each other. By Theorem 1,
\begin{equation*}
\lim_{p_0\rightarrow\infty}\mathrm{FDP}(t)\stackrel{\mathrm{D}}{=}\frac{p_0\Big[\Phi(d(z_{t/2}+\sqrt{\rho}W))+\Phi(d(z_{t/2}-\sqrt{\rho}W))\Big]}{\sum_{i=1}^p\Big[\Phi(d(z_{t/2}+\sqrt{\rho}W+\mu_i))+\Phi(d(z_{t/2}-\sqrt{\rho}W-\mu_i))\Big]},
\end{equation*}
where $d=(1-\rho)^{-1/2}$.

\textbf{Example 2: [Multifactor Model]} Consider a multifactor model:
\begin{equation}\label{c4}
Z_i=\mu_i+\eta_i+a_i^{-1}K_i, \quad\quad i=1,\cdots,p,
\end{equation}
where $\eta_i$ and $a_i$ are defined in Theorem 1 and $K_i\sim N(0,1)$ for $i=1,\cdots,p$. All the $W_h$'s and $K_i$'s are independent of each other. In this model, $W_1,\cdots,W_k$ are the $k$ common factors. By Theorem 1, expression (\ref{a50}) holds.

Although the distribution of $\FDP(t)$ in Example 2 is the same as that in Theorem 1, the extension from Example 2 to Theorem 1 is technical. The key difference is that Example 2 assumes a multifactor model with independent random errors for the test statistics and Theorem 1 relaxes this restricted assumption largely to the arbitrary dependence structure.

In Theorem 1, since $\FDP$ is bounded by 1, taking expectation on both sides of the equation (\ref{a50}) and by the Portmanteau lemma, we have the convergence of FDR:
\begin{corollary}
Under the assumptions in Theorem 1, for the high dimensional sparse case,
\begin{equation}\label{a51}
\lim_{p_0\rightarrow\infty}\mathrm{FDR}(t)=E\Big[\frac{\sum_{i\in\text{\{true null\}}}\Big\{\Phi(a_i(z_{t/2}+\eta_i))+\Phi(a_i(z_{t/2}-\eta_i))\Big\}}{\sum_{i=1}^p\Big\{\Phi(a_i(z_{t/2}+\eta_i+\mu_i))+\Phi(a_i(z_{t/2}-\eta_i-\mu_i))\Big\}}\Big].
\end{equation}
\end{corollary}
The expectation on the right hand side of (\ref{a51}) is with respect to standard multivariate normal variables $(W_1,\cdots,W_k)^T\sim N_k(0,\bI_k)$.

The proof of Theorem 1 is based on the following result.
\begin{proposition}
Under the assumptions in Theorem 1,
\begin{eqnarray}
\lim_{p\rightarrow\infty}p^{-1}R(t)
\stackrel{D}{=}p^{-1}\sum_{i=1}^p\Big[\Phi(a_i(z_{t/2}+\eta_i+\mu_i))+\Phi(a_i(z_{t/2}-\eta_i-\mu_i))\Big], \label{a54}\\
\lim_{p_0\rightarrow\infty}p_0^{-1}V(t)
\stackrel{D}{=}p_0^{-1}\sum_{i\in\text{\{true null\}}}\Big[\Phi(a_i(z_{t/2}+\eta_i))+\Phi(a_i(z_{t/2}-\eta_i))\Big].
\end{eqnarray}
\end{proposition}
The proofs of Theorem 1 and Proposition 2 are shown in the Appendix.

\subsection{Estimating FDP}
In Theorem 1 and Proposition 2, the summation over the set of true null hypotheses is uncomputable, because it is not known which factor loadings correspond to the true nulls. However, due to the high dimensionality and sparsity, both $p$ and $p_0$ are large and $p_1$ is relatively small. Therefore, we can use
\begin{equation}\label{a52}
\sum_{i=1}^p\Big[\Phi(a_i(z_{t/2}+\eta_i))+\Phi(a_i(z_{t/2}-\eta_i))\Big]
\end{equation}
as a conservative surrogate for
\begin{equation}\label{a53}
\sum_{i\in\text{\{true null\}}}\Big[\Phi(a_i(z_{t/2}+\eta_i))+\Phi(a_i(z_{t/2}-\eta_i))\Big].
\end{equation}
Since only $p_1$ extra terms are included in (\ref{a52}), the substitution is accurate enough.

The mean of $V(t)$ is $E\Big[\sum_{i\in\text{\{true null\}}}I(P_i\leq t)\Big]=p_0t$, since the $P$-values corresponding to the true null hypotheses are uniformly distributed. However, the variance of $V(t)$ can be large when the test statistics $Z_1,\cdots,Z_p$ are dependent. Owen (2005) has theoretically studied the variance of the number of false discoveries. In our framework, expression (\ref{a52}) is a function of i.i.d. standard normal variables. Given $t$, the variance of (\ref{a52}) can be obtained by simulations and hence variance of $V(t)$ is approximated via (\ref{a52}). Relevant simulation studies will be presented in Section 5.

In recent years, there have been substantial interests in the realized random variable FDP itself, instead of controlling FDR, as we are usually concerned about the number of false discoveries in a given experiment, rather than an average of FDP for hypothetical replications of the experiment. See Genovese \& Wasserman (2004), Meinshausen (2005), Efron (2007), etc. In our problem, it is known that the approximate asymptotic distribution of $V(t)/p_0$ is
\begin{equation}
p_0^{-1}\sum_{i=1}^p\Big[\Phi(a_i(z_{t/2}+\eta_i))+\Phi(a_i(z_{t/2}-\eta_i))\Big].
\end{equation}
Let
\begin{equation*}
\mathrm{FDP_A}(t)=\Big(\sum_{i=1}^p\Big[\Phi(a_i(z_{t/2}+\eta_i))+\Phi(a_i(z_{t/2}-\eta_i))\Big]\Big)/R(t),
\end{equation*}
if $R(t)\neq0$ and $\mathrm{FDP_A}(t)=0$ when $R(t)=0$. Given observations $z_1,\cdots,z_p$ of the test statistics $Z_1,\cdots,Z_p$, if the unobserved but realized factors $W_1,\cdots,W_k$ can be estimated by $\widehat{W}_1,\cdots,\widehat{W}_k$, then we can obtain an estimator of $\mathrm{FDP_A}(t)$ by
\begin{equation} \label{b21}
\widehat{\FDP}(t)=\min\Big(\sum_{i=1}^p\Big[\Phi(a_i(z_{t/2}+\widehat{\eta}_i))+\Phi(a_i(z_{t/2}-\widehat{\eta}_i))\Big],R(t)\Big)/R(t),
\end{equation}
when $R(t)\neq0$ and $\widehat{\FDP}(t)=0$ when $R(t)=0$. Note that in (\ref{b21}), $\widehat{\eta}_i=\sum_{h=1}^kb_{ih}\widehat{W}_h$ is an estimator for $\eta_i=\sum_{h=1}^kb_{ih}W_h$.

The following procedure is one practical way to estimate $W_1,\cdots,W_k$ based on the data. For observed values $z_1,\cdots,z_p$, we choose the smallest $75\%$ of $|z_i|$'s. For ease of notation, assume the first $m$ $z_i$'s have the smallest absolute values. Then approximately
\begin{equation} \label{b22}
Z_i=\sum_{h=1}^kb_{ih}W_h+K_i,\quad i=1,\cdots,m.
\end{equation}
The approximation from (\ref{b20}) to (\ref{b22}) stems from the intuition that large $|\mu_i|$'s tend to produce large $|z_i|$'s and the sparsity makes approximation errors negligible. Finally we apply the robust $L_1$-regression to the equation set (\ref{b22}) and obtain the least-absolute deviation estimates $\widehat{W}_1,\cdots,\widehat{W}_k$. The estimator (\ref{b21}) performs significantly better than Efron (2007)'s estimator in our simulation studies. One difference is that in our setting $\bSigma$ is known. The other is that we give a better approximation as shown in Section 3.4.

\subsection{Asymptotic Justification}
Theorem 2 shows the asymptotic consistency of $L_1-$regression estimators under model (\ref{b22}). Portnoy (1984b) has proven the asymptotic consistency for robust regression estimation when the random errors are i.i.d. However, his proof does not work here because of the weak dependence of  random errors. Our result allows $k$ to grow with $m$, even at a faster rate of $o(m^{1/4})$ imposed by Portnoy (1984b).
\begin{theorem}
Suppose (\ref{b22}) is a correct model. Let $\hw$ be the $L_1-$regression estimator:
\begin{equation}
\hw \equiv \argmin_{\bbeta\in R^k}\sum_{i=1}^m|Z_i-\bb_i^T\bbeta|
\end{equation}
where $\bb_i=(b_{i1},\cdots,b_{ik})^T$. Let $\bw=(w_1,\cdots,w_k)^T$ be the realized values of $\{W_h\}_{h=1}^k$.
Suppose $k=O(m^{\kappa})$ for $0\leq\kappa<1-\delta$. Under the assumptions
\begin{itemize}
\item[(C1)]\label{a55}
$\sum_{j=k+1}^p\lambda_j^2\leq\eta$ for $\eta=O(m^{2\kappa})$,
\item[(C2)]\label{a56}
There exists a constant $d>0$ such that
\begin{equation*}
\lim_{m\rightarrow\infty}\sup_{\|\bu\|=1}m^{-1}\sum_{i=1}^mI(|\bb_i^T\bu|\leq d)=0,
\end{equation*}
\item[(C3)] $a_{\max}/a_{\min}\leq S$ for some constant $S$ when $m\rightarrow\infty$ where $1/a_i$ is the standard deviation of $K_i$,
\item[(C4)] $a_{\min}=O(m^{(1-\kappa)/2})$.
\end{itemize}
We have $\|\hw-\bw\|_2=O_p(\sqrt{\frac{k}{m}})$.
\end{theorem}
(C1) is stronger than (C0) in Theorem 1 as (C0) only requires $\sum_{j=k+1}^p\lambda_j^2=O(p^{2-2\delta})$. (C2) ensures the identifiability of $\bbeta$, which is similar to Proposition 3.3 in Portnoy (1984a). (C3) and (C4) are imposed to facilitate the technical proof.

We now show in Theorem 3 the asymptotic consistency of $\widehat{\FDP}(t)$ based on $L_1-$regression estimator of $\bW=(W_1,\cdots,W_k)^T$ in model (\ref{b22}). The proof of Theorem 3 is based on the result in Theorem 2.
\begin{theorem}
If the assumptions in Theorem 2 are satisfied, and in addition, the following conditions are satisfied:
\begin{itemize}
\item[(C5)] $R(t)/p>H$ for $H>0$ as $p\rightarrow\infty$,
\item[(C6)] $\min_{1\leq i\leq p}\min(|z_{t/2}+\bb_i^T\bw|,|z_{t/2}-\bb_i^T\bw|)\geq \tau>0$,
\end{itemize}
then $|\mathrm{\widehat{FDP}}(t)-\mathrm{FDP_A}(t)|=O_p(\sqrt{\frac{k}{m}})$.
\end{theorem}
In Theorem 3, (C6) is a reasonable condition because $z_{t/2}$ is a negative number when threshold $t$ is very small and $\bb_i^T\bw$ is a realization from a normal distribution $N(0, \sum_{h=1}^kb_{ih}^2)$ with $\sum_{h=1}^kb_{ih}^2<1$. Thus $z_{t/2}+\bb_i^T\bw$ or $z_{t/2}-\bb_i^T\bw$ is unlikely close to zero. Our proof shows a more general result. Under conditions of Theorem 3,
\begin{equation*}
|\mathrm{\widehat{FDP}}(t)-\mathrm{FDP_A}(t)|=O_p(\|\hw-\bw\|_2)
\end{equation*}
\indent The results in Theorems 2--3 are based on the assumption that (\ref{b22}) is a correct model. In the following we will show that even if (\ref{b22}) is not a correct model, the effects of misspecification are negligible when $p$ is sufficiently large. To facilitate the mathematical derivations, we instead consider the least-squares estimator. Suppose we are estimating $\bW=(W_1,\cdots,W_k)^T$ from (\ref{b20}). Without loss of generality, assume the true values of $\{W_h\}_{h=1}^p$ are 0. Let $\bX$ be the design matrix of model (\ref{b20}), then the least-squares estimator for $\bW$ is $\widehat{\bW}_{\LS}^*=(\bX^T\bX)^{-1}\bX^T(\bmu+\bK)$, where $\bmu=(\mu_1,\cdots,\mu_p)^T$ and $\bK=(K_1,\cdots,K_p)^T$. Instead, we estimate $W_1,\cdots,W_k$ based on the simplified model (\ref{b22}) with $m=p$, which ignores sparse $\{\mu_i\}$, then the least-squares estimator for $\bW$ is $\widehat{\bW}_{\LS}=(\bX^T\bX)^{-1}\bX^T\bK$. The following result shows that the effect of misspecification in model (\ref{b22}) is negligible when $p\rightarrow\infty$:
\begin{theorem}
The bias due to ignoring non-nulls is controlled by
\begin{equation*}
\|\widehat{\bW}_{\mathrm{LS}}-\widehat{\bW}_{\mathrm{LS}}^*\|_2\leq\|\bmu\|_2\Big(\sum_{i=1}^k\lambda_i^{-1}\Big)^{1/2}
\end{equation*}
\end{theorem}

In Theorem 1, we can choose appropriate $k$ such that $\lambda_k>1$ and $\lambda_i\rightarrow\infty$ as $p\rightarrow\infty$ for $i\leq k$. Therefore, $\sum_{i=1}^k\lambda_i^{-1}\rightarrow 0$ as $p\rightarrow\infty$ is a reasonable condition. When $\{\mu_i\}_{i=1}^p$ are truly sparse, it is expected that $\|\bmu\|_2$ grows slowly or is even bounded so that the bound in Theorem 4 is small. For $L_1-$regression, it is expected to be even more robust to the outliers in the sparse vector $\{\mu_i\}_{i=1}^p$.

\subsection{Relation with Other Methods}
Efron (2007) proposed a novel parametric model for $V(t)$:
\begin{equation}
V(t)=p_0t\Big[1+2A\frac{(-z_{t/2})\phi(z_{t/2})}{\sqrt{2}t}\Big],
\end{equation}
where $A\sim N(0,\alpha^2)$ for some real number $\alpha$ and $\phi(\cdot)$ stands for the probability density function of standard normal distribution. The correlation effect is explained by the dispersion variate $A$. His procedure is to estimate $A$ from the data and use
\begin{equation}
p_0t\Big[1+2\widehat{A}\frac{(-z_{t/2})\phi(z_{t/2})}{\sqrt{2}t}\Big]\Big/R(t)
\end{equation}
as an estimator for FDP$(t)$. Note that the above expressions are adaptations from his procedure for the one-sided test to our two-sided test setting. In his simulation, the above estimator captures the general trend of the FDP, but it is not accurate and deviates from the true FDP with large amount of noise. Consider our estimator $\widehat{\mathrm{FDP}}(t)$ in (\ref{b21}). Write $\widehat{\eta}_i=\sigma_iQ_i$ where $Q_i\sim N(0,1)$. When $\sigma_i\rightarrow0$ for $\forall i\in\{\text{true null}\}$, by the second order Taylor expansion,
\begin{equation}
\widehat{\mathrm{FDP}}(t)\approx\frac{p_0t}{R(t)}\Big[1+\sum_{i\in\{\text{true null}\}}\sigma_i^2(Q_i^2-1)\frac{(-z_{t/2})\phi(z_{t/2})}{p_0t}\Big].
\end{equation}
By comparison with Efron's estimator, we can see that
\begin{equation}
\widehat{A}=\frac{1}{\sqrt{2}p_0}\sum_{i\in\{\text{true null}\}}\Big[\widehat{\eta}_i^2-E(\widehat{\eta}_i^2)\Big].
\end{equation}
Thus, our method is more general and more precise.

Leek \& Storey (2008) considered a general framework for modeling the dependence in multiple testing. Their idea is to model the dependence via a factor model and reduces the multiple testing problem from dependence to independence case via accounting the effects of common factors. They also provided a method of estimating the common factors. In contrast, our problem is different from Leek \& Storey's and we estimate common factors from very different methods. In addition, we provide the approximated FDP formula and its consistent estimate.

Friguet, Kloareg \& Causeur (2009) followed closely the framework of Leek \& Storey (2008). They assumed that the data come directly from a multifactor model with independent random errors, and then used the EM algorithm to estimate all the parameters in the model and obtained an estimator for FDP$(t)$. In particular, they subtract $\eta_i$ out of (\ref{c4}) based on their estiamte from the EM algorithm to improve the efficiency. However, the estimated number of factors in their studies is usually small by their EM algorithm, thus leading to inaccurate estimated $\FDP(t)$. Moreover, it is hard to derive theoretical results based on the estimator from their EM algorithm. Compared with their results, our procedure does not assume any specific dependence structure of the test statistics. What we do is to decompose the test statistics into an approximate factor model with weakly dependent errors, derive the factor loadings and estimate the unobserved but realized factors by $L_1$-regression. Since the theoretical distribution of $V(t)$ is known, estimator (\ref{b21}) performs well based on a good estimation for $W_1,\cdots,W_k$. See Section 5 for relevant simulation results.

\section{Approximate Control of FDR}
In this section we will propose some ideas that can asymptotically control the FDR, not the FDP, under arbitrary dependency. Although their validity is yet to be established, promising results reveal in the simulation studies. Therefore, they are worth some discussion and serve as a direction of our future work.

Suppose the number of false null hypotheses $p_1$ is known. If the signal $\mu_i$ for $i\in\text{\{false null\}}$ is strong enough such that
\begin{equation} \label{b24}
\Phi\Big(a_i(z_{t/2}+\eta_i+\mu_i)\Big)+\Phi\Big(a_i(z_{t/2}-\eta_i-\mu_i)\Big)\approx1,
\end{equation}
then asymptotically the FDR is approximately given by
\begin{equation} \label{b25}
\FDR(t)=E\Big\{\frac{\sum_{i=1}^p\Big[\Phi(a_i(z_{t/2}+\eta_i))+\Phi(a_i(z_{t/2}-\eta_i))\Big]}{\sum_{i=1}^p\Big[\Phi(a_i(z_{t/2}+\eta_i))+\Phi(a_i(z_{t/2}-\eta_i))\Big]+p_1}\Big\},
\end{equation}
which is the expectation of a function of $W_1,\cdots,W_k$. Note that $\FDR(t)$ is a known function and can be computed by Monte Carlo simulation. For any predetermined error rate $\alpha$, we can use the bisection method to solve $t$ so that $\FDR(t)=\alpha$. Since $k$ is not large, the Monte Carlo computation is sufficiently fast for most applications.

The requirement (\ref{b24}) is not very strong. First of all, $\Phi(3)\approx0.9987$, so (\ref{b24}) will hold if any number inside the $\Phi(\cdot)$ is greater than 3. Secondly, $1-\sum_{h=1}^kb_{ih}^2$ is usually very small. For example, if it is $0.01$, then $a_i=(1-\sum_{h=1}^kb_{ih}^2)^{-1/2}\approx10$, which means that if either $z_{t/2}+\eta_i+\mu_i$ or $z_{t/2}-\eta_i-\mu_i$ exceed 0.3, then (\ref{b24}) is approximately satisfied. Since the effect of sample size $n$ is involved, (\ref{b24}) is not a very strong condition on the signal strength $\{\beta_i\}$.

\section{Simulation Studies}
In the simulation studies, we consider $p=2000$, $n=100$, $\sigma=2$, the number of false null hypotheses $p_1=10$ and the nonzero $\beta_i=1$, unless stated otherwise. We will present 6 different dependence structures for $\bSigma$ of the test statistics $(Z_1,\cdots,Z_p)^T\sim N((\mu_1,\cdots,\mu_p)^T,\bSigma)$. Following the setting in section 2, $\bSigma$ is the correlation matrix of a random sample of size $n$ of $p-$dimensional vector $\bX_i=(X_{i1},\cdots,X_{ip})$, and $\mu_j=\sqrt{n}\beta_j\widehat{\sigma}_j/\sigma$, $j=1,\cdots,p$. The data generating process vector $\bX_i$'s are as follows.
\begin{itemize}
\item \textbf{[Equal correlation]} Let $\bX^T=(X_{1},\cdots,X_{p})^T\sim N_p(0,\bSigma)$ where $\bSigma$ has diagonal element 1 and off-diagonal element $1/2$.
\item \textbf{[Fan \& Song's model]} For $\bX=(X_{1},\cdots,X_{p})$, let $\{X_{k}\}_{k=1}^{1900}$ be i.i.d. $N(0,1)$ and
    \begin{equation*}
    X_{k}=\sum_{l=1}^{10}X_{l}(-1)^{l+1}/5+\sqrt{1-\frac{10}{25}}\epsilon_{k}, \ \ k=1901,\cdots,2000,
    \end{equation*}
    where $\{\epsilon_{k}\}_{k=1901}^{2000}$ are standard normally distributed.
\item \textbf{[Independent Cauchy]} For $\bX=(X_{1},\cdots,X_{p})$, let $\{X_{k}\}_{k=1}^{2000}$ be i.i.d. Cauchy random variables with location parameter 0 and scale parameter 1.
\item \textbf{[Three factor model]} For $\bX=(X_{1},\cdots,X_{p})$, let
    \begin{equation*}
    X_{j}=\rho_j^{(1)}W^{(1)}+\rho_j^{(2)}W^{(2)}+\rho_j^{(3)}W^{(3)}+H_{j},
    \end{equation*}
    where $W^{(1)}\sim N(-2,1)$, $W^{(2)}\sim N(1,1)$, $W^{(3)}\sim N(4,1)$, $\rho_{j}^{(1)},\rho_{j}^{(2)},\rho_{j}^{(3)}$ are i.i.d. $U(-1,1)$, and $H_{j}$ are i.i.d. $N(0,1)$.
\item \textbf{[Two factor model]} For $\bX=(X_{1},\cdots,X_{p})$, let
    \begin{equation*}
    X_{j}=\rho_j^{(1)}W^{(1)}+\rho_j^{(2)}W^{(2)}+H_{j},
    \end{equation*}
    where $W^{(1)}$ and $W^{(2)}$ are i.i.d. $N(0,1)$, $\rho_{j}^{(1)}$ and $\rho_{j}^{(2)}$ are i.i.d. $U(-1,1)$, and $H_{j}$ are i.i.d. $N(0,1)$.
\item \textbf{[Nonlinear factor model]} For $\bX=(X_{1},\cdots,X_{p})$, let
    \begin{equation*}
    X_{j}=\sin(\rho_j^{(1)}W^{(1)})+sgn(\rho_j^{(2)})\exp(|\rho_j^{(2)}|W^{(2)})+H_{j},
    \end{equation*}
    where $W^{(1)}$ and $W^{(2)}$ are i.i.d. $N(0,1)$, $\rho_{j}^{(1)}$ and $\rho_{j}^{(2)}$ are i.i.d. $U(-1,1)$, and $H_{j}$ are i.i.d. $N(0,1)$.
\end{itemize}

Fan \& Song's Model has been considered in Fan \& Song (2010) for high dimensional variable selection. This model is close to the independent case but has some special dependence structure. Note that although we have used the term ``factor model" above to describe the dependence structure, it is not the factor model for the test statistics $Z_1,\cdots,Z_p$ directly. Instead, we assume a factor model for the data $X_1,\cdots,X_p$ to construct some other dependence structures for the covariance matrix $\bSigma$.

\begin{figure}[h!!!]
\setlength{\unitlength}{1mm}
\begin{center}
\scalebox{0.38}{\includegraphics{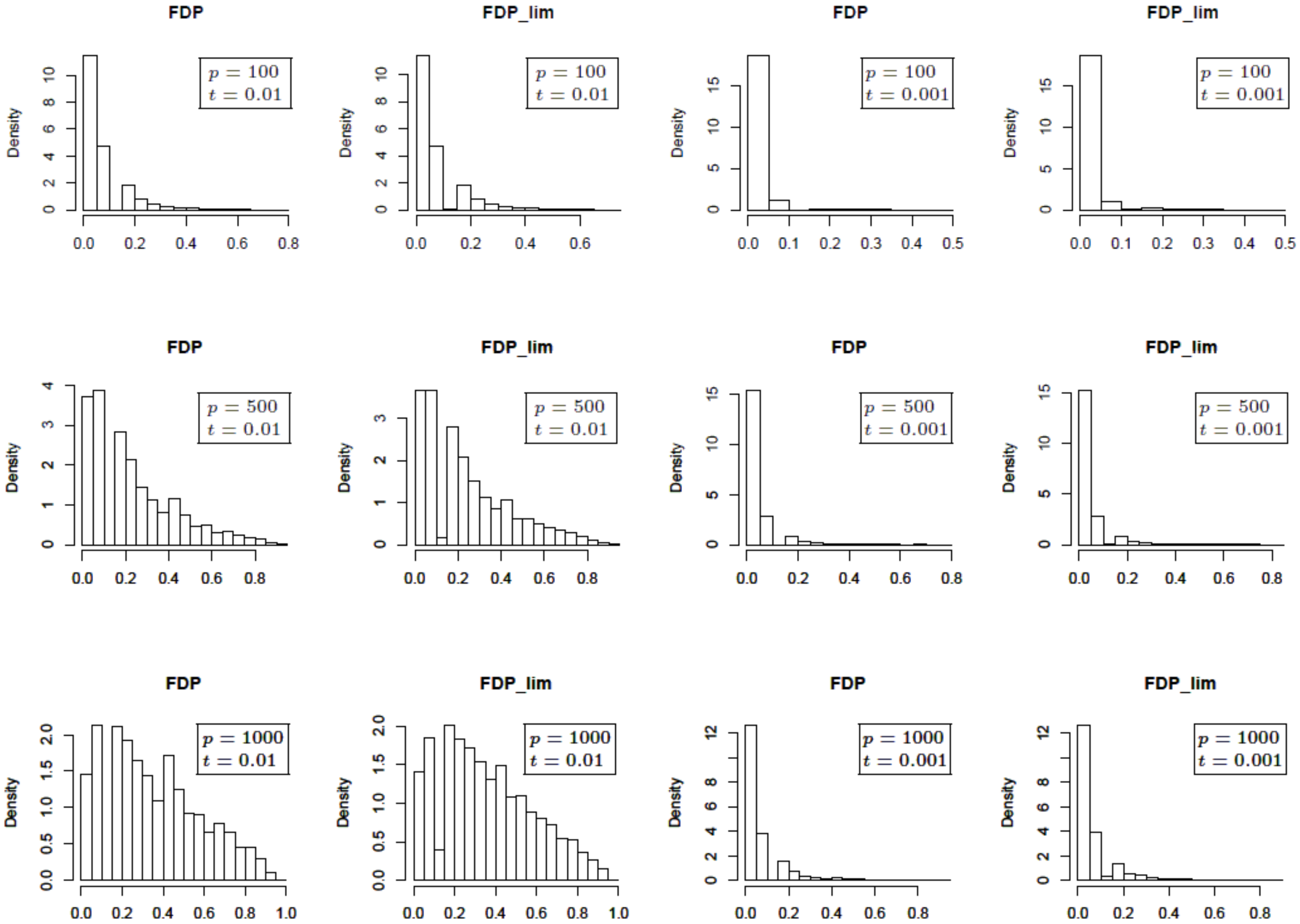}}
\end{center}
\caption{Comparison for the distribution of the FDP with the limiting distribution of the FDP, based on the two factor model over 10000 simulations. From the top row to the bottom, $p=100, 500, 1000$ respectively. The first two columns correspond to $t=0.01$ and the last two correspond to $t=0.001$. The first and the third columns are for the FDP, while the second and the fourth are for the limit of the FDP.}\label{ar}
\label{a59}
\end{figure}

\textbf{Convergence of FDP:} Our result in Theorem 1 is based on asymptotic convergence. Without loss of generality, we consider a dependence structure based on the two factor model above. Let $n=100$, $p_1=10$ and $\sigma=2$. Let $p$ vary from 100 to 1000 and $t$ be either 0.01 or 0.001. In Figure~\ref{a59}, we will show that the convergence is fast as $p$ increases. Therefore, the asymptotic result in Theorem 1 should work very well when there are hundreds or thousands of hypotheses tested simultaneously.

\textbf{Variance of $V(t)$:} Variance of false discoveries in the correlated test statistics is usually large compared with that of the independent case, due to correlation structures. In Table 2, for high dimensional sparse case, we compare the true variance of number of false discoveries, the variance of expression (\ref{a53}) (which is infeasible in practice) and the variance of expression (\ref{a52}) under 6 different dependence structures. It shows that the variance computed based on expression (\ref{a52}) approximately equals the variance of number of false discoveries. Therefore for high dimensional sparse case, we provide a fast and alternative method to estimate the variance of number of false discoveries in addition to the results in Owen (2005).

\begin{table}[h!!!]
\begin{center}\caption{Comparison for variance of number of false discoveries (column 2), variance of expression (\ref{a53}) (column 3) and variance of expression (\ref{a52}) (column 4) with $t=0.001$ based on 10000 simulations. }\label{ar}

  \begin{tabular}{crrr}
  \hline\hline
   Dependence Structure  & $\var(V(t))$  & $\var(V)$  & $\var(V.up)$\\
   \hline
     Equal correlation   &  180.9673     &  178.5939  & 180.6155    \\
     Fan \& Song's model &   5.2487     & 5.2032   &   5.2461  \\
     Independent Cauchy  &  9.0846     & 8.8182   &   8.9316  \\
     Three factor model  &   81.1915   & 81.9373    & 83.0818  \\
     Two factor model    &  53.9515    & 53.6883    &  54.0297   \\
     Nonlinear factor model  &  48.3414    & 48.7013    &   49.1645  \\
   \hline
   \end{tabular}
\end{center}
\end{table}

\begin{table}[h!!!]
\begin{center}\caption{Comparison of FDR values for our method based on equation (\ref{b25}) (PFA) with Storey's procedure and Benjamini-Hochberg's procedure under six different dependence structures, where $p=2000$, $n=200$, $t=0.001$, and $\beta_i=1$ for $i\in\text{\{false null\}}$. The computation is based on 10000 simulations and Monte Carlo errors are listed in the brackets.}\label{e1}
  \begin{tabular}{ccccc}
  \hline\hline
                         & True FDR  & PFA  & Storey   & B-H\\
   \hline
     Equal correlation   & $6.67\%$     & $6.61\%$   & $2.99\%$  &  $3.90\%$   \\
                         & ($15.87\%$)  &($15.88\%$) &($10.53\%$)& ($14.58\%$) \\
     Fan \& Song's model & $14.85\%$     & $14.85\%$    & $13.27\%$   & $14.46\%$   \\
                         & ($11.76\%$)   &($11.58\%$)   &($11.21\%$)  &($13.46\%$)  \\
     Independent Cauchy  & $13.85\%$     & $13.62\%$  & $11.48\%$  & $13.21\%$  \\
                         &($13.60\%$)    &($13.15\%$) &($12.39\%$) &($15.40\%$) \\
     Three factor model  & $8.08\%$      & $8.29\%$    & $4.00\%$   & $5.46\%$  \\
                         &($16.31\%$)   &($16.39\%$)   & ($11.10\%$)& ($16.10\%$)\\
     Two factor model    & $8.62\%$     & $8.50\%$     & $4.70\%$ & $5.87\%$    \\
                         &($16.44\%$)   &($16.27\%$)   &($11.97\%$) &($16.55\%$) \\
     Nonlinear factor model  & $6.63\%$     & $6.81\%$     & $3.20\%$ & $4.19\%$    \\
                         & ($15.56\%$)    & ($15.94\%$)    &($10.91\%$) &($15.31\%$)\\
   \hline
   \end{tabular}
\end{center}
\end{table}

\textbf{Comparing methods of controlling FDR:} Under different dependence structures, we compare FDR values for our procedure PFA in equation (\ref{b25}) with $p_1$ known, Storey's procedure and Benjamini-Hochberg procedure. Table~\ref{e1} shows that our method performs much better than Storey's procedure and Benjamini-Hochberg procedure, especially under strong dependence structures (rows 1, 4, 5, and 6), in terms of both mean and variance of the distribution of FDP.
\begin{figure}[h!!!]
\begin{center}
\scalebox{0.43}{\includegraphics{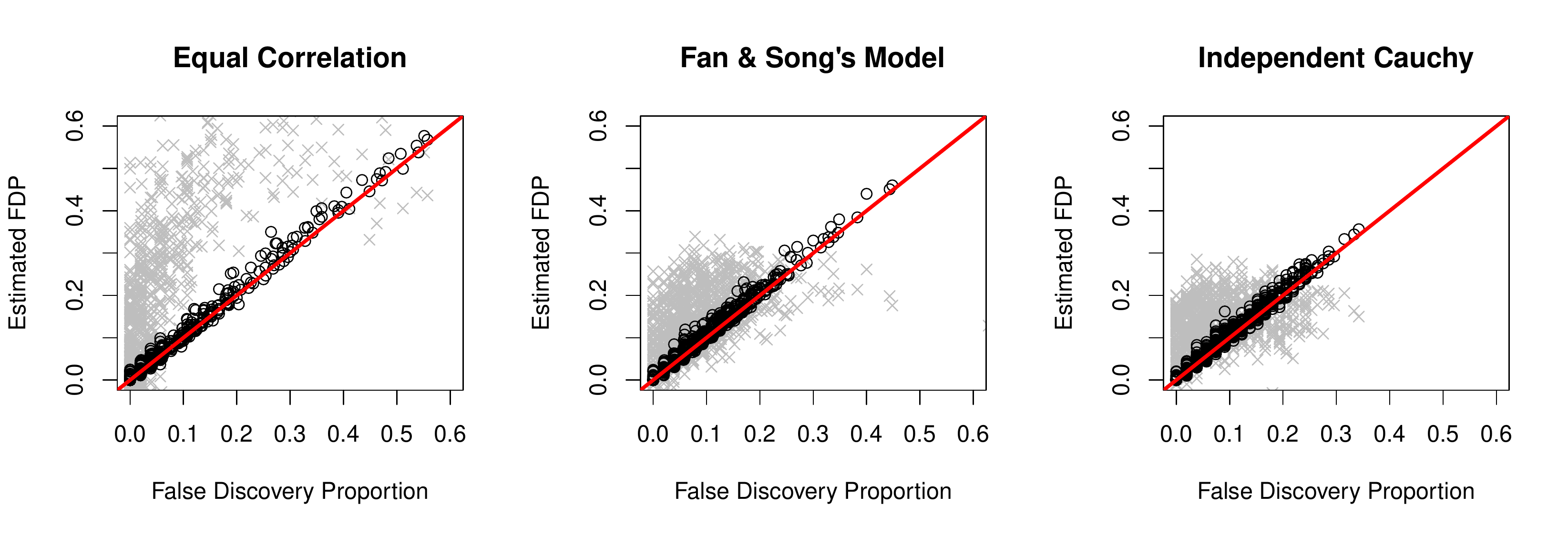}}
\scalebox{0.43}{\includegraphics{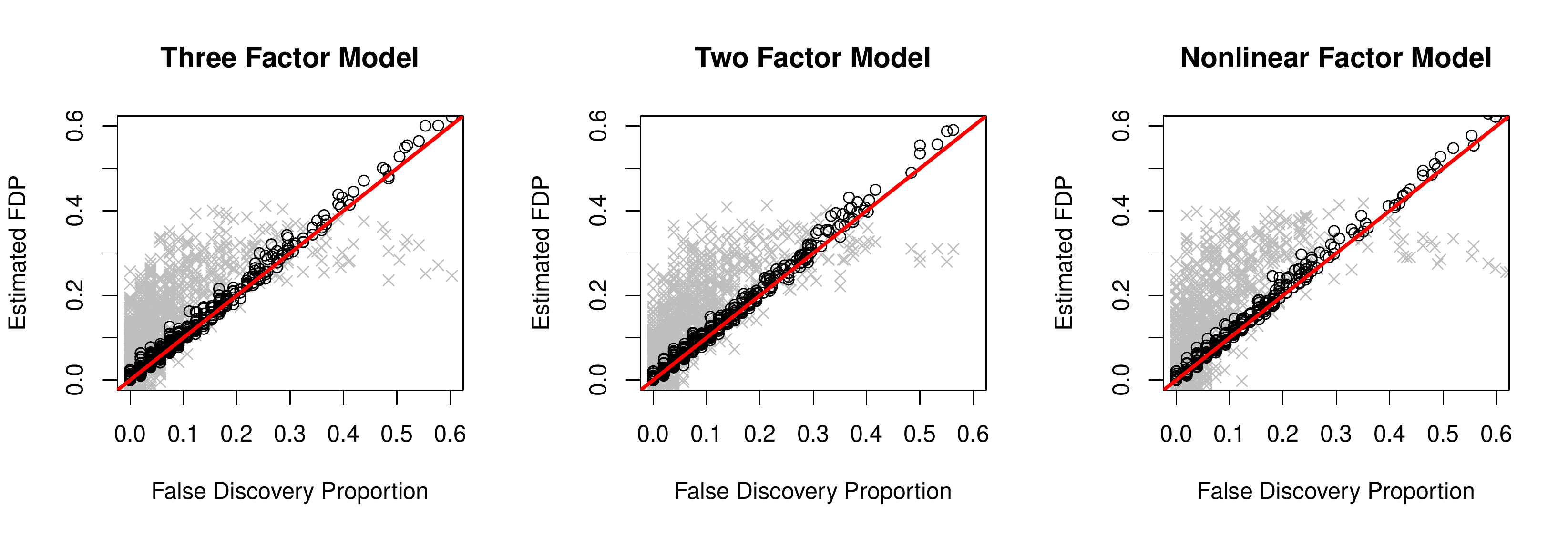}}
\end{center}
\vspace{-0.3cm}
\caption{Comparison of true values of False Discovery Proportion with estimated FDP by Efron (2007)'s procedure (crosses) and our PFA method (dots) under six different dependence structures, with $p=1000$, $p_1=50$, $n=100$, $\sigma=2$, $t=0.005$ and $\beta_i=1$ for $i\in\text{\{false null\}}$ based on 1000 simulations. The $Z$-statistics with absolute value less than or equal to $x_0=1$ are used to estimate the dispersion variate $A$ in Efron (2007)'s estimator.}
\label{a60}
\end{figure}

\begin{figure}[h!!!]
\begin{center}
\scalebox{0.43}{\includegraphics{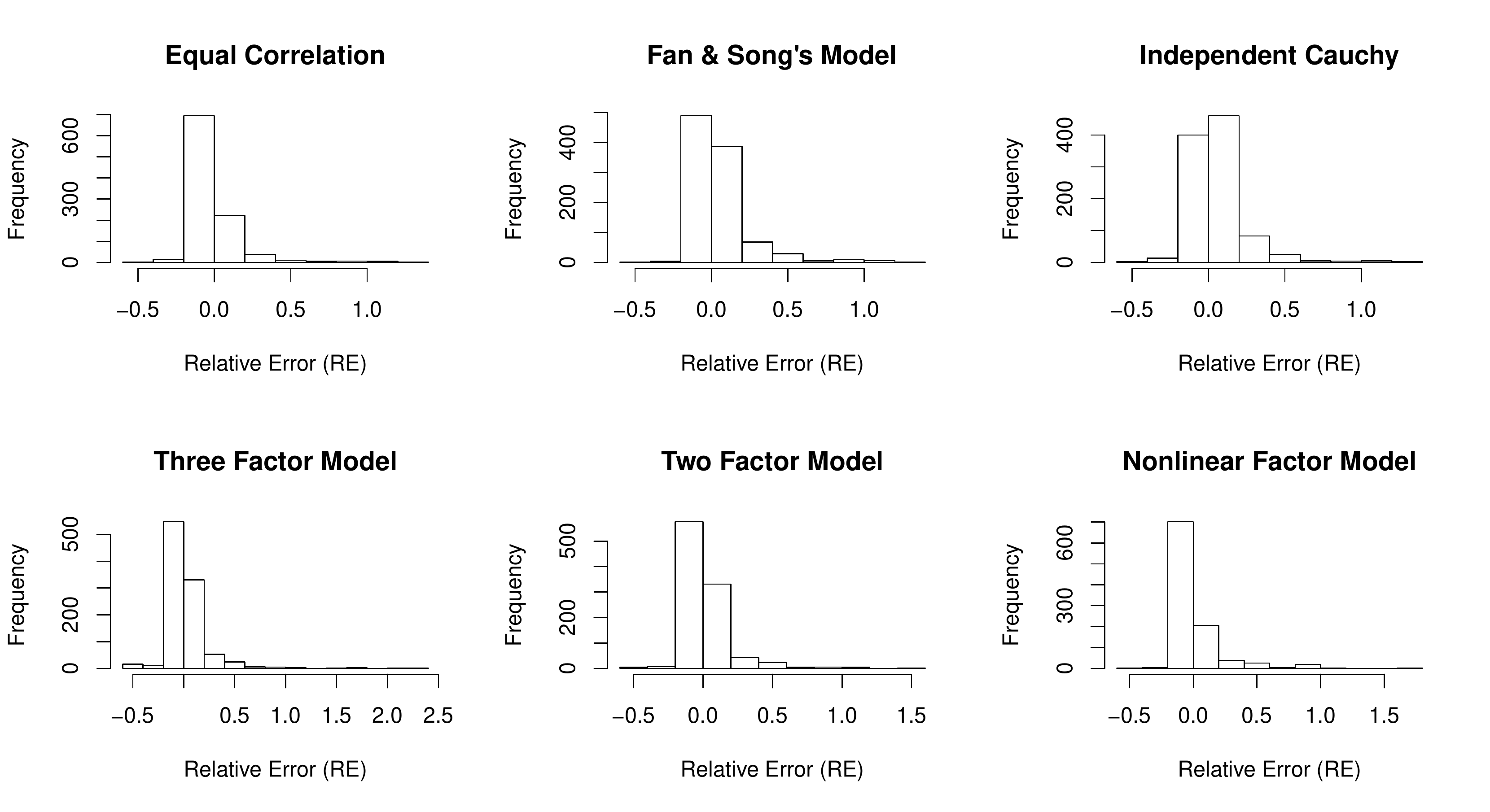}}
\end{center}
\vspace{-0.3cm}
\caption{Histograms of the relative error (RE) between true values of FDP and estimated FDP by our PFA method under the six dependence structures in Figure~\ref{a60}. RE is defined as $(\widehat{\text{FDP}}(t)-\text{FDP}(t))/\text{FDP}(t)$ if $\text{FDP}(t) \neq 0$ and 0 otherwise.}\label{c7}
\end{figure}

\textbf{Estimate FDP:} We now compare the estimated values of our method PFA (\ref{b21}) and Efron (2007)'s estimator with true values of false discovery proportion, under 6 different dependence structures. The results are depicted in Figure~\ref{a60}, Figure~\ref{c7} and Table~\ref{e2}. Figure~\ref{a60} shows that our estimated values correctly track the trends of FDP with smaller amount of noise. It also shows that both estimators tend to overestimate the true FDP, since $\mathrm{FDP_A}(t)$ is an upper bound of the true $\mathrm{FDP}(t)$. They are close only when the number of false nulls $p_1$ is very small. In the current simulation setting, we choose $p_1=50$ compared with $p=1000$, therefore, it is not a very sparse case. However, even under this case, our estimator still performs very well for six different dependence structures. Efron (2007)'s estimator is illustrated in Figure~\ref{a60} with his suggestions for estimating parameters, which captures the general trend of true FDP but with large amount of noise.

\begin{table}[h!]
\begin{center}\caption{Means and standard deviations of the relative error between true values of FDP and estimated FDP under the six dependence structures in Figure~\ref{a60}. $\text{RE}_{\text{P}}$ is the relative error of our PFA estimator and $\text{RE}_{\text{E}}$ is the relative error of Efron (2007)'s estimator. RE is defined in Figure~\ref{c7}.}\label{e2}
  \begin{tabular}{ccccc}
  \hline\hline
                   & mean($\text{RE}_{\text{P}}$)  & SD($\text{RE}_{\text{P}}$)   & mean($\text{RE}_{\text{E}}$)   & SD($\text{RE}_{\text{E}}$)   \\
   \hline
     Equal correlation   & 0.0342   & 0.1579   & 1.6132     &3.5320 \\
     Fan \& Song's model & 0.0685   & 0.1801   & 1.1549     &1.7180 \\
     Independent Cauchy  & 0.0611   & 0.1685   & 1.3086     &2.0900 \\
     Three factor model  & 0.0456   & 0.2072   & 1.1953     &2.3070 \\
     Two factor model    & 0.0428   & 0.1625   & 1.0658     &1.9039 \\
     Nonlinear factor model     &  0.0548   &  0.1815  &  1.2446    &2.4927 \\
   \hline
   \end{tabular}
\end{center}
\end{table}

\section{Real Data Analysis}
Our proposed multiple testing procedures are now applied to the genome-wide association studies, in particular the expression quantitative trait locus (eQTL) mapping. It is known that the expression levels of gene CCT8 are highly related to Down Syndrome phenotypes. In our analysis, we use over two million SNP genotype data and CCT8 gene expression data for 210 individuals from three different populations, testing which SNPs are associated with the variation in CCT8 expression levels. To save space, we omit the description of the data pre-processing procedures. Interested readers can find more details from the websites: http://pngu.mgh.harvard.edu/~purcell/plink/res.shtml and ftp://ftp.sanger.ac.uk/pub/genevar/, and the paper Bradic, Fan \& Wang (2010).

We further introduce two sets of dummy variables $(\mbox{\bf d}_1,\mbox{\bf d}_2)$ to recode the SNP data, where $\mbox{\bf d}_1=(d_{1,1},\cdots,d_{1,p})$ and $\mbox{\bf d}_2=(d_{2,1},\cdots,d_{2,p})$, representing three categories of polymorphisms, namely, $(d_{1,j},d_{2,j})=(0,0)$ for $\text{SNP}_j=0$ (no polymorphism), $(d_{1,j},d_{2,j})=(1,0)$ for $\text{SNP}_j=1$ (one nucleotide has polymorphism) and $(d_{1,j},d_{2,j})=(0,1)$ for $\text{SNP}_j=2$ (both nucleotides have polymorphisms). Thus, instead of using model (\ref{gwj1}), we consider two marginal linear regression models between $Y$ and $d_{1,j}$:
\begin{equation}\label{gwj2}
\min_{\alpha_{1,j},\beta_{1,j}}E(Y-\alpha_{1,j}-\beta_{1,j} d_{1,j})^2, \ \ \ j=1,\cdots,p
\end{equation}
and between $Y$ and $d_{2,j}$:
\begin{equation}\label{gwj3}
\min_{\alpha_{2,j},\beta_{2,j}}E(Y-\alpha_{2,j}-\beta_{2,j} d_{2,j})^2, \ \ \ j=1,\cdots,p.
\end{equation}
For ease of notation, we denote the recoded $n \times 2p$ dimensional design matrix as $\bX$. The missing SNP measurement are imputed as $0$ and the redundant SNP data are excluded. Finally, logarithm-transform of the raw CCT8 gene expression data are used.
The details of our testing procedures are summarized as follows.
\begin{itemize}
\item
To begin with, consider the full model $Y=\alpha+\bX\beta + \epsilon$, where $Y$ is the CCT8 gene expression data, $\bX$ is the $n \times 2p$ dimensional design matrix of the SNP codings and $\epsilon_i\sim N(0,\sigma^2)$, $i=1,\cdots,n$ are the independent random errors. We adopt the refitted cross-validation (RCV) (Fan, Guo \& Hao 2010) technique to estimate $\sigma$ by $\widehat{\sigma}$, where LASSO is used in the first (variable selection) stage.
\item
Fit the marginal linear models (\ref{gwj2}) and (\ref{gwj3}) for each (recoded) SNP and obtain the least-squares estimate $\widehat{\beta}_j$ for $j=1,\cdots,2p$. Compute the values of $Z$-statistics using formula (\ref{b2}), except that $\sigma$ is replaced by $\widehat{\sigma}$.
\item
Calculate the P-values based on the $Z$-statistics and compute $R(t)=\#\{P_j: P_j \leq t\}$ for a fixed threshold $t$.
\item
Apply eigenvalue decomposition to the covariance matrix $\bSigma$ of the $Z$-statistics. Determine an appropriate number of factors $k$ and derive the corresponding factor loading coefficients $\{b_{ih}\}_{i=1,\ h=1}^{i=2p,\ h=k}$.
\item
Order the absolute-valued $Z$-statistics and choose the first $m=95\%\times2p$ of them. Apply $L_1$-regression to the equation set (\ref{b22}) and obtain its solution $\widehat{W}_1,\cdots,\widehat{W}_k$. Plug them into (\ref{b21}) and get the estimated $\text{FDP}(t)$.
\end{itemize}

\begin{figure}[h!!!]
\begin{center}
\scalebox{0.44}[0.40]{\includegraphics{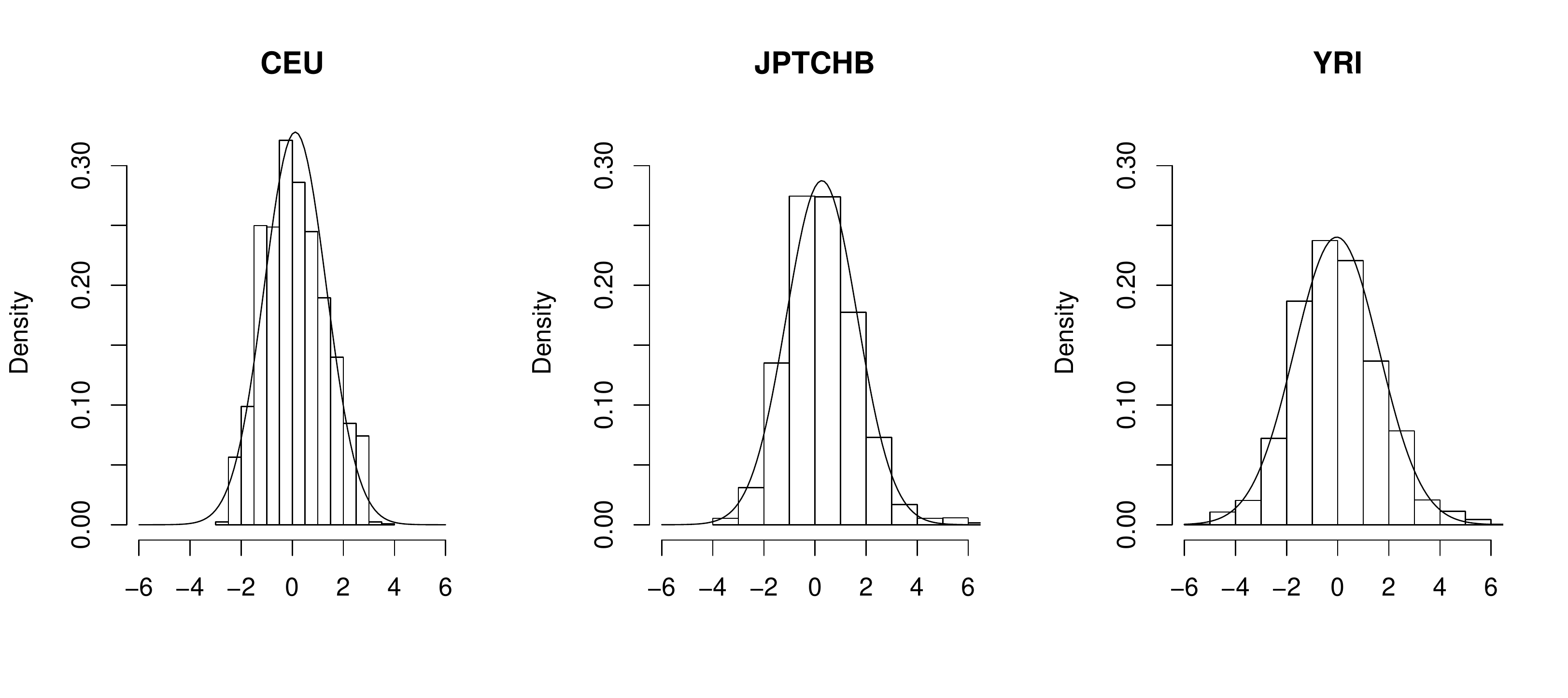}}
\end{center}
\vspace{-0.3cm}
\caption{Empirical distributions and fitted normal density curves of the $Z$-values for each of the three populations. Because of dependency, the $Z$-values are no longer $N(0,1)$ distributed. The empirical distributions, instead, are $N(0.12,1.22^2)$ for CEU, $N(0.27,1.39^2)$ for JPT and CHB, and $N(-0.04,1.66^2)$ for YRI, respectively. The density curve for CEU is closest to $N(0,1)$ and the least dispersed among the three.}
\label{a62}
\end{figure}

\begin{figure}[h!!!]
\begin{center}
\scalebox{0.44}[0.40]{\includegraphics{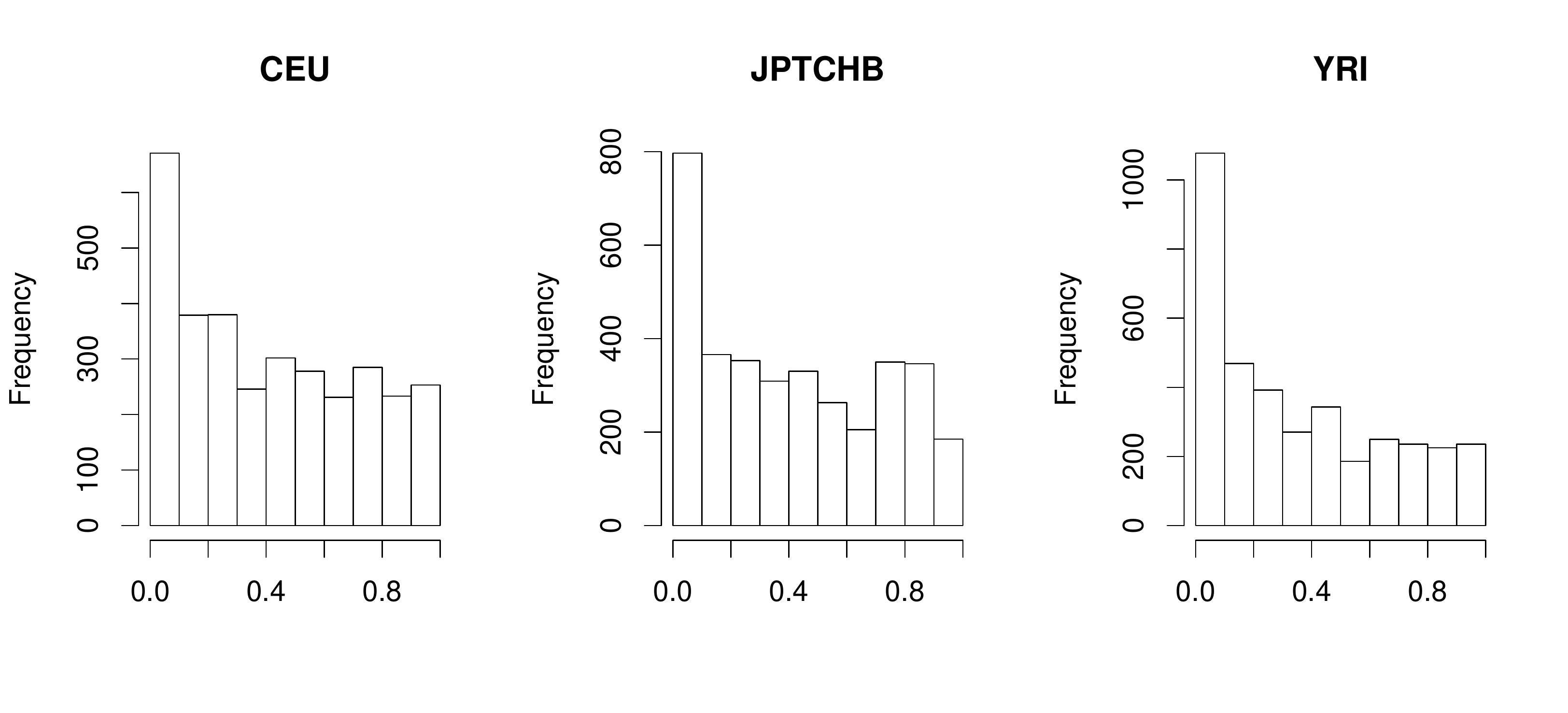}}
\end{center}
\vspace{-0.3cm}
\caption{Histograms of the $P$-values for each of the three populations. }
\label{a63}
\end{figure}

For each intermediate step of the above procedure, the outcomes are summarized in the following figures. Figure~\ref{a61} illustrates the trend of the RCV-estimated standard deviation $\widehat{\sigma}$ with respect to different model sizes. Our result is similar to that in Fan, Guo \& Hao (2010), in that although $\widehat{\sigma}$ is influenced by the selected model size, it is relatively stable and thus provides reasonable accuracy. The empirical distributions of the $Z$-values are presented in Figure~\ref{a62}, together with the fitted normal density curves. As pointed out in Efron (2007, 2010), due to the existence of dependency among the $Z$-values, their densities are either narrowed or widened and are not $N(0,1)$ distributed. The histograms of the $P$-values are further provided in Figure~\ref{a63}, giving a crude estimate of the proportion of the false nulls for each of the three populations.

\begin{figure}
\begin{center}
\scalebox{0.44}[0.40]{\includegraphics{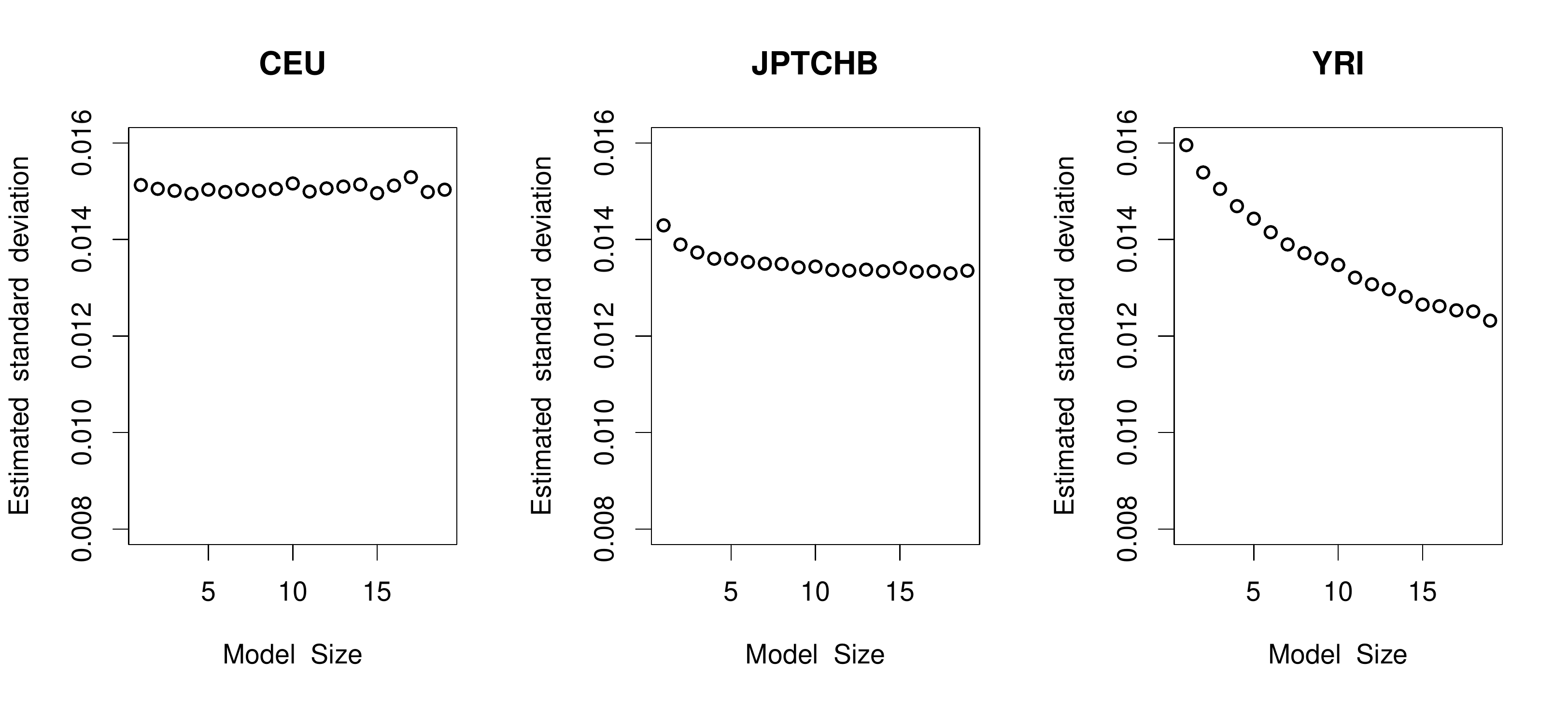}}
\end{center}
\vspace{-0.3cm}
\caption{$\widehat{\sigma}$ of the three populations with respect to the selected model sizes, derived by using refitted cross-validation (RCV).}
\label{a61}
\end{figure}

\begin{figure}[h!!!]
\begin{center}
\scalebox{0.43}[0.40]{\includegraphics{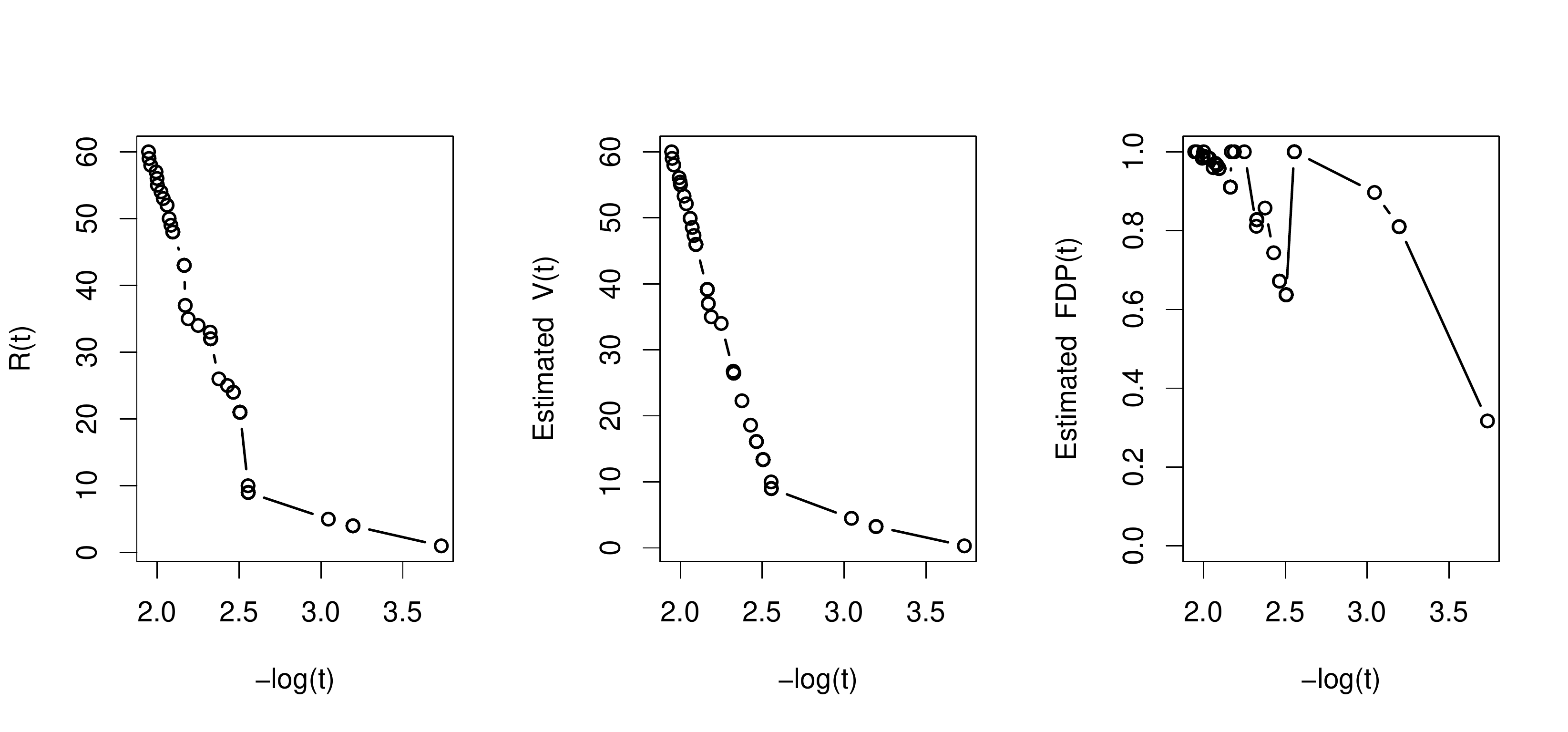}}
\scalebox{0.43}[0.40]{\includegraphics{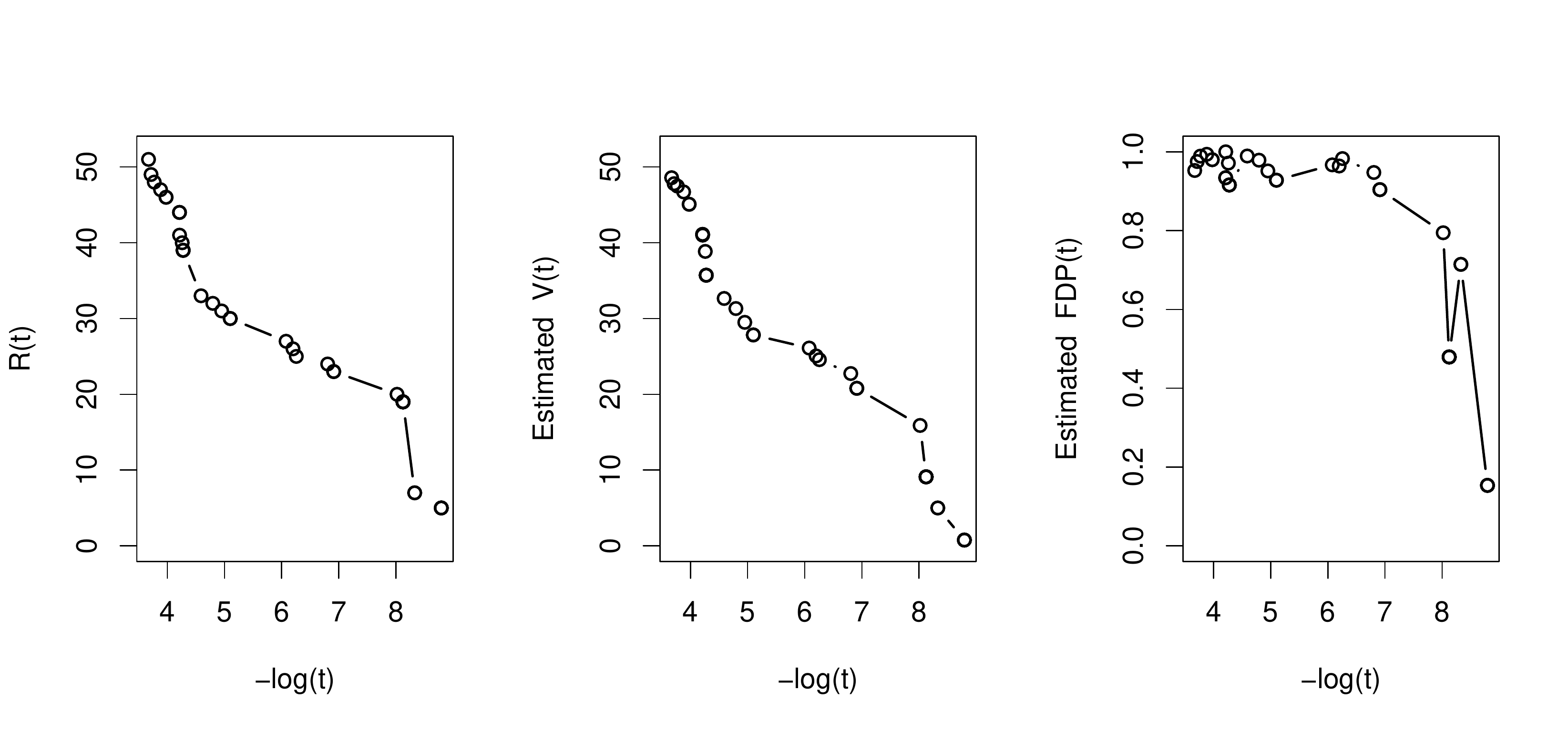}}
\scalebox{0.43}[0.40]{\includegraphics{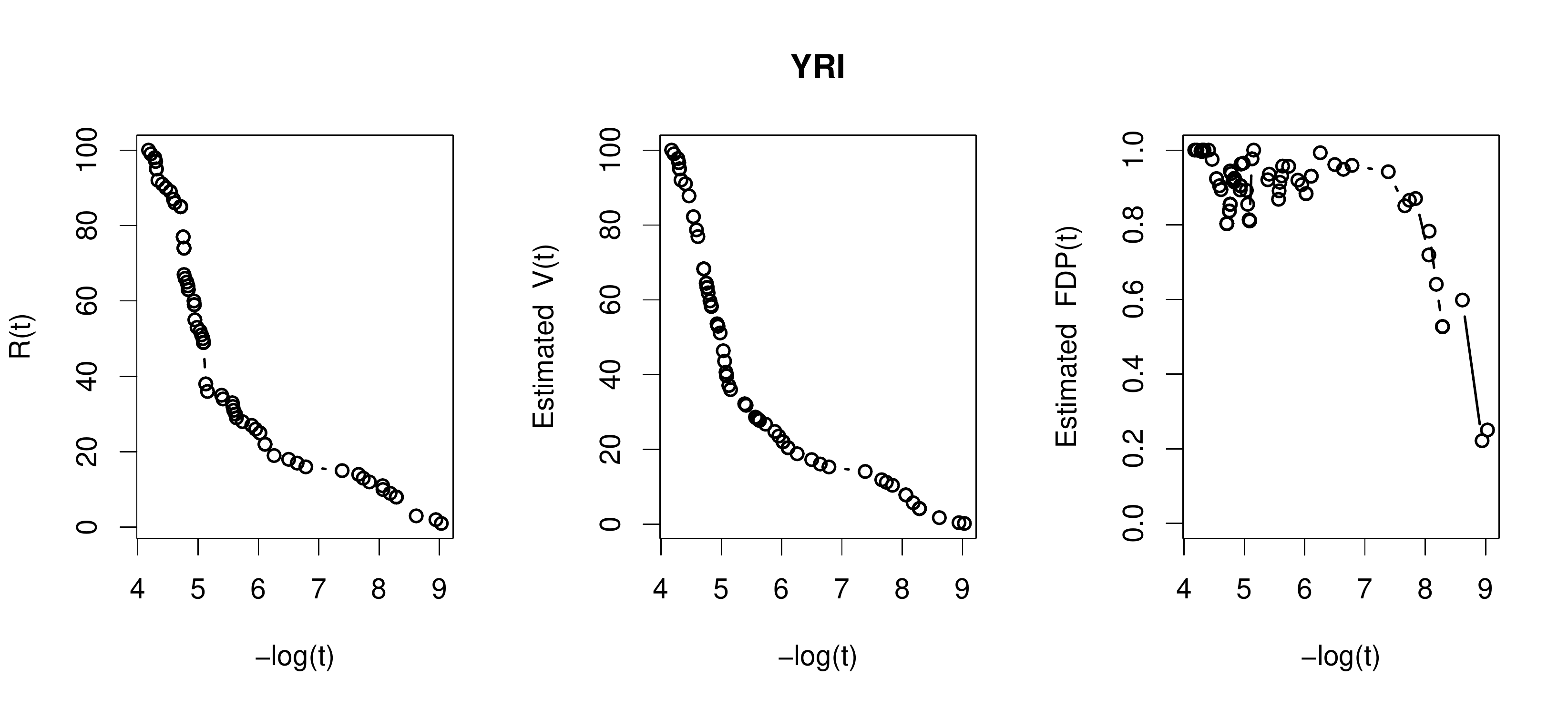}}
\end{center}
\vspace{-0.3cm}
\caption{Number of total discoveries, estimated number of false discoveries and estimated False Discovery Proportion as functions of thresholding $t$ for CEU population (row 1), JPT and CHB (row 2) and YRI (row 3). The $x$-coordinate is $-\log t$, the minus $\log_{10}$-transformed thresholding.}\label{a64}
\end{figure}

The main results of our analysis are presented in Figures~\ref{a64}, which depicts the number of total discoveries $R(t)$, the estimated number of false discoveries $\widehat{V}(t)$ and the estimated False Discovery Proportion $\widehat{\text{FDP}}(t)$ as functions of (the minus $\log_{10}$-transformed) thresholding $t$ for the three populations. As can be seen, in each case both $R(t)$ and $\widehat{V}(t)$ are decreasing when $t$ decreases, but $\widehat{\text{FDP}}(t)$ exhibits zigzag patterns and does not always decrease along with $t$, which results from the cluster effect of the P-values. A closer study of the outputs further shows that for all populations, the estimated FDP has a general trend of decreasing to the limit of around $0.1$ to $0.2$, which backs up the intuition that a large proportion of the smallest $P$-values should correspond to the false nulls (true discoveries) when Z-statistics is very large; however, in most other thresholding values, the estimated FDPs are at a high level. This is possibly due to small signal-to-noise ratios in eQTL studies.

The results of the selected SNPs, together with the estimated FDPs, are depicted in Table~\ref{a67}. It is worth mentioning that Deutsch et al. (2005) and Bradic, Fan \& Wang (2010) had also worked on the same CCT8 data to identify the significant SNPs in CEU population. Deutsch et al. (2005) performed association analysis for each SNP using ANOVA, while Bradic, Fan \& Wang (2010) proposed the penalized composite quasi-likelihood variable selection method. Their findings were different as well, for the first group identified four SNPs (exactly the same as ours) which have the smallest P-values but the second group only discovered one SNP rs965951 among those four, arguing that the other three SNPs make little additional contributions conditioning on the presence of rs965951. Our results for CEU population coincide with that of the latter group, in the sense that the false discovery rate is high in our findings and our association study is marginal rather than joint modeling among several SNPs.

\begin{table}[h!]
\begin{center}\caption{Information of the selected SNPs and the associated FDP for a particular threshold. Note that the density curve of the $Z$-values for CEU population is close to $N(0,1)$, so the approximate $\widehat{\text{FDP}}(t)$ equals $pt/R(t)\approx 0.631$. Therefore our high estimated FDP is reasonable.}\label{a67}
  \begin{tabular}{c c c c l}
  \hline\hline
   Population  & Threshold  & \# Discoveries  &  Estimated FDP  &  Selected SNPs \\
   \hline
     JPTCHB   & $1.61 \times 10^{-9}$   & 5  & $  0.1535$   &  rs965951 rs2070611\\
              &                         &    &              &  rs2832159 rs8133819\\
              &                         &    &              &  rs2832160 \\
   \hline
     YRI      & $1.14 \times 10^{-9}$   & 2  & $  0.2215$   &  rs9985076 rs965951\\
   \hline
     CEU      & $6.38 \times 10^{-4}$   & 4  & $  0.8099$   &  rs965951 rs2832159 \\
              &                         &    &              &  rs8133819 rs2832160\\
   \hline
   \end{tabular}

\end{center}
\end{table}

\section{Discussion}
We have proposed a new method (principal factor approximation) for high dimensional multiple testing where the test statistics have an arbitrary dependence structure. For multivariate normal test statistics with a known covariance matrix, we can express the test statistics as an approximate factor model with weakly dependent random errors, by applying eigenvalue decomposition to the covariance matrix. We show the theoretical distribution of the false discovery proportion in large scale simultaneous tests when a common threshold is used for rejection. This result has important applications in controlling FDP and FDR. We also provide a procedure to estimate the realized FDP, which, in our simulation studies, correctly tracks the trend of FDP with smaller amount of noise.

In the current paper, a fixed threshold is used for multiple testing under arbitrary dependency. Our future research interests will focus on how to take advantage of the dependence structure such that the testing procedure is more powerful or even optimal under arbitrary dependence structures. One possible way is to vary, according to the dependence structure, the threshold values for different hypotheses, based on successive conditioning.

\section{Appendix}
Lemma 1 is fundamental to our proof of Theorem 1 and Proposition 2. The result is known in probability, but has the formal statement and proof in Lyons (1988).
\begin{lemma}[Strong Law of Large Numbers for Weakly Correlated Variables]
Let $\{X_n\}_{n=1}^{\infty}$ be a sequence of real-valued random variables such that $E|X_n|^2\leq1$. If $|X_n|\leq1$ a.s. and $\sum_{N\geq1}\frac{1}{N}E|\frac{1}{N}\sum_{n\leq N}X_n|^2<\infty$, then $\lim_{N\rightarrow\infty}\frac{1}{N}\sum_{n\leq N}X_n=0 \ \ a.s.$.
\end{lemma}
\noindent\textbf{Proof of Proposition 2:} Note that $P_i = 2\Phi(-|Z_i|)$.
Based on the expression of $(Z_1,\cdots,Z_p)^T$ in (\ref{b20}), $\Big\{I(P_i\leq t|W_1,\cdots,W_k)\Big\}_{i=1}^p$ are dependent random variables. Nevertheless, we want to prove
\begin{equation} \label{b27}
p^{-1}\sum_{i=1}^pI(P_i\leq t|W_1,\cdots,W_k)\stackrel{p\rightarrow\infty}{\longrightarrow}p^{-1}\sum_{i=1}^pP(P_i\leq t|W_1,\cdots,W_k) \ a.s. .
\end{equation}
Letting $X_i=I(P_i\leq t|W_1,\cdots,W_k)-P(P_i\leq t|W_1,\cdots,W_k)$, by Lemma 1 the conclusion (\ref{b27}) is correct if we can show
\begin{equation*}
\Var\Big(p^{-1}\sum_{i=1}^{p}I(P_i\leq t|W_1,\cdots,W_k)\Big)=O_p(p^{-\delta}) \ \ \text{for some} \ \delta>0.
\end{equation*}
\vspace{-0.1cm}
To begin with, note that
\begin{eqnarray*}
&&\Var\Big(p^{-1}\sum_{i=1}^{p}I(P_i\leq t|W_1,\cdots,W_k)\Big)\\
&=&p^{-2}\sum_{i=1}^{p}\Var\Big(I(P_i\leq t|W_1,\cdots,W_k)\Big)\\
&&+2p^{-2}\sum_{1\leq i<j\leq p}\Cov\Big(I(P_i\leq t|W_1,\cdots,W_k),I(P_j\leq t|W_1,\cdots,W_k)\Big).
\end{eqnarray*}
Since $\Var\big(I(P_i\leq t|W_1,\cdots,W_k)\big)\leq\frac{1}{4}$, the first term in the right-hand side of the last equation is $O_p(p^{-1})$. For the second term, the covariance is given by
\begin{equation*}
P(P_i\leq t,P_j\leq t|W_1,\cdots,W_k)-P(P_i\leq t|W_1,\cdots,W_k)P(P_j\leq t|W_1,\cdots,W_k).
\end{equation*}
To simplify the notation, let $\rho_{ij}^k$ be the correlation between $K_i$ and $K_j$. Without loss of generality, we assume $\rho_{ij}^k>0$ (for $\rho_{ij}^k<0$, the calculation is similar). Denote by
\begin{equation*}
c_{1,i}= a_i(-z_{t/2}-\eta_i-\mu_i), \ \ \ c_{2,i}= a_i(z_{t/2}-\eta_i-\mu_i).
\end{equation*}
Then, from the joint normality, it can be shown that
\begin{eqnarray}\label{e3}
&&P(P_i\leq t,P_j\leq t|W_1,\cdots,W_k)\nonumber\\
&=&P(c_{2,i}/a_i<K_i<c_{1,i}/a_i, c_{2,j}/a_j<K_j<c_{1,j}/a_j)\nonumber\\
&=&\int_{-\infty}^{\infty}\Big[\Phi\Big(\frac{(\rho_{ij}^k)^{1/2}z+c_{1,i}}{(1-\rho_{ij}^k)^{1/2}}\Big)-\Phi\Big(\frac{(\rho_{ij}^k)^{1/2}z+c_{2,i}}{(1-\rho_{ij}^k)^{1/2}}\Big)\Big]\\
&&\quad\quad\times\Big[\Phi\Big(\frac{(\rho_{ij}^k)^{1/2}z+c_{1,j}}{(1-\rho_{ij}^k)^{1/2}}\Big)-\Phi\Big(\frac{(\rho_{ij}^k)^{1/2}z+c_{2,j}}{(1-\rho_{ij}^k)^{1/2}}\Big)\Big]\phi(z)dz.\nonumber
\end{eqnarray}

Next we will use Taylor expansion to analyze the joint probability further. We have shown that $(K_1,\cdots,K_p)^T\sim N(0,\bA)$ are weakly dependent random variables. Let $cov_{ij}^k$ denote the covariance of $K_i$ and $K_j$, which is the $(i,j)$th element of the covariance matrix $\bA$. We also let $b_{ij}^k=(1-\sum_{h=1}^kb_{ih}^2)^{1/2}(1-\sum_{h=1}^kb_{jh}^2)^{1/2}$. By the H\"{o}lder inequality,
\begin{eqnarray*}
p^{-2}\sum_{i,j=1}^p|cov_{ij}^k|^{1/2}\leq p^{-1/2}(\sum_{i,j=1}^p|cov_{ij}^k|^2)^{1/4}=\Big[p^{-2}(\sum_{i=k+1}^{p}\lambda_i^2)^{1/2}\Big]^{1/4}\rightarrow0
\end{eqnarray*}
as $p\rightarrow\infty$. For each $\Phi(\cdot)$, we apply Taylor expansion with respect to $(cov_{ij}^k)^{1/2}$,
\begin{eqnarray*}
\Phi\Big(\frac{(\rho_{ij}^k)^{1/2}z+c_{1,i}}{(1-\rho_{ij}^k)^{1/2}}\Big)&=&\Phi\Big(\frac{(cov_{ij}^k)^{1/2}z+(b_{ij}^k)^{1/2}c_{1,i}}{(b_{ij}^k-cov_{ij}^k)^{1/2}}\Big)
\end{eqnarray*}
\begin{eqnarray*}
      &=&\Phi(c_{1,i})+\phi(c_{1,i})(b_{ij}^k)^{-1/2}z(cov_{ij}^k)^{1/2}\\
      &&\quad\quad\quad+\frac{1}{2}\phi(c_{1,i})c_{1,i}(b_{ij}^k)^{-1}(1-z^2)cov_{ij}^k+o(cov_{ij}^k).
\end{eqnarray*}
Therefore, we have (\ref{e3}) equals
\begin{eqnarray*}
&&\Big[\Phi(c_{1,i})-\Phi(c_{2,i})\Big]\Big[\Phi(c_{1,j})-\Phi(c_{2,j})\Big]\\
&&\quad\quad+\Big(\phi(c_{1,i})-\phi(c_{2,i})\Big)\Big(\phi(c_{1,j})-\phi(c_{2,j})\Big)(b_{ij}^k)^{-1}cov_{ij}^k+o(cov_{ij}^k),
\end{eqnarray*}
where we have used the fact that $\int_{-\infty}^{\infty} z\phi(z)dz=0$ and $\int_{-\infty}^{\infty} (1-z^2)\phi(z)dz=0$. Now since $P(P_i\leq t|W_1,\cdots,W_k)=\Phi(c_{1,i})-\Phi(c_{2,i})$, we have
\begin{eqnarray*}
&&\Cov\Big(I(P_i\leq t|W_1,\cdots,W_k),I(P_j\leq t|W_1,\cdots,W_k)\Big)\\
&=&\Big(\phi(c_{1,i})-\phi(c_{2,i})\Big)\Big(\phi(c_{1,j})-\phi(c_{2,j})\Big)a_ia_jcov_{ij}^k+o(cov_{ij}^k).
\end{eqnarray*}
In the last line, $\big(\phi(c_{1,i})-\phi(c_{2,i})\big)\big(\phi(c_{1,j})-\phi(c_{2,j})\big)a_ia_j$ is bounded by some constant except on a countable collection of measure zero sets. Let $C_i$ be defined as the set $\{z_{t/2}+\eta_i+\mu_i=0\}\cup\{z_{t/2}-\eta_i-\mu_i=0\}$. On the set $C_i^c$, $\big(\phi(c_{1,i})-\phi(c_{2,i})\big)a_i$ converges to zero as $a_i\rightarrow\infty$. Therefore, $\big(\phi(c_{1,i})-\phi(c_{2,i})\big)\big(\phi(c_{1,j})-\phi(c_{2,j})\big)a_ia_j$ is bounded by some constant on $(\bigcup_{i=1}^pC_i)^c$.

By the Cauchy-Schwartz inequality and $(C0)$ in Theorem 1, $p^{-2}\sum_{i,j}|cov_{i,j}^k|=O(p^{-\delta})$. On the set $(\bigcup_{i=1}^pC_i)^c$, we conclude that

\begin{equation*}
\Var\Big(p^{-1}\sum_{i=1}^pI(P_i\leq t|W_1,\cdots,W_k)\Big)=O_p(p^{-\delta}).
\end{equation*}
Hence by Lemma 1,
\begin{equation*}
p^{-1}\sum_{i=1}^pI(P_i\leq t|W_1,\cdots,W_k)\stackrel{p\rightarrow\infty}{\longrightarrow}p^{-1}\sum_{i=1}^pP(P_i\leq t|W_1,\cdots,W_k) \ a.s..
\end{equation*}
Therefore,
\begin{eqnarray*}
\lim_{p\rightarrow\infty}p^{-1}\sum_{i=1}^pI(P_i\leq t)
&\stackrel{D}{=}&p^{-1}\sum_{i=1}^pP(P_i\leq t|W_1,\cdots,W_k)\\
&=&p^{-1}\sum_{i=1}^p\Big[\Phi(a_i(z_{t/2}+\eta_i+\mu_i))+\Phi(a_i(z_{t/2}-\eta_i-\mu_i))\Big].
\end{eqnarray*}

With the same argument we can also show
\begin{equation*}
\lim_{p_0\rightarrow\infty}p_0^{-1}V(t)
\stackrel{D}{=}p_0^{-1}\sum_{i\in\text{\{true null\}}}\Big[\Phi(a_i(z_{t/2}+\eta_i))+\Phi(a_i(z_{t/2}-\eta_i))\Big]
\end{equation*}
for the high dimensional sparse case. The proof of Proposition 2 is now complete.

\noindent\textbf{Proof of Theorem 1:}\\
For all bounded continuous functions $f$,
\begin{eqnarray*}
\lim_{p_0\rightarrow\infty}E\Big[f\Big(\frac{V(t)}{R(t)}\Big)\Big]
&=&E\Big[E\Big[f\Big(\lim_{p_0\rightarrow\infty}\frac{(p_0/p)(V(t)/p_0)}{R(t)/p}\Big)|W_1,\cdots,W_k\Big]\Big]\\
&=&E\Big[E\Big[f\Big(\frac{(p_0/p)(\sum_{i\in\text{\{true  null\}}}P(P_i\leq t|W_1,\cdots,W_k)/p_0)}{\sum_{i=1}^pP(P_i\leq t|W_1,\cdots,W_k)/p}\Big)\Big]\Big]\\
&=&E\Big[f\Big(\frac{\sum_{i\in\text{\{true  null\}}}P(P_i\leq t|W_1,\cdots,W_k)}{\sum_{i=1}^pP(P_i\leq t|W_1,\cdots,W_k)}\Big)\Big].
\end{eqnarray*}
In the second equality, we have used the convergence results in Proposition 2 and the Continuous Mapping Theorem. We have also used the Slutsky's Theorem, since, given $W_1,\cdots,W_k$, $R(t)/p$ converges in distribution to a constant $p^{-1}\sum_{i=1}^pP(P_i\leq t|W_1,\cdots,W_k)$. Finally, by the Portmanteau's Lemma,
\begin{eqnarray*}
\lim_{p_0\rightarrow\infty}\frac{V(t)}{R(t)}
&\stackrel{D}{=}&\frac{\sum_{i\in\text{\{true  null\}}}P(P_i\leq t|W_1,\cdots,W_k)}{\sum_{i=1}^pP(P_i\leq t|W_1,\cdots,W_k)}\\
&=& \frac{\sum_{i\in\text{\{true null\}}}\Big[\Phi(a_i(z_{t/2}+\eta_i))+\Phi(a_i(z_{t/2}-\eta_i))\Big]}{\sum_{i=1}^p\Big[\Phi(a_i(z_{t/2}+\eta_i+\mu_i))+\Phi(a_i(z_{t/2}-\eta_i-\mu_i))\Big]}.
\end{eqnarray*}
The proof of Theorem 1 is complete.

\noindent\textbf{Proof of Theorem 2:}  Without loss of generality, we assume that the true value of $\bw$ is zero, and we need to prove $\|\hw\|_2=O_p(\sqrt{\frac{k}{m}})$. Let $L: R^m\rightarrow R^m$ be defined by
\begin{equation*}
L_j(\bw)=m^{-1}\sum_{i=1}^m b_{ij}sgn(K_i-\bb_i^T\bw)
\end{equation*}
where $sgn(x)$ is the sign function of $x$ and equals zero when $x=0$. Then we want to prove that there is a root $\hw$ of the equation $L(\bw)=0$ satisfying $\|\hw\|_2^2=O_p(k/m)$. We will apply Result 6.3.4 of Ortega and Rheinboldt (1970, page 163), so it suffices to show that with high probability, $\bw^TL(\bw)<0$ with $\|\bw\|_2^2=Bk/m$ for a sufficiently large constant $B$.

Let $V=\bw^TL(\bw)=m^{-1}\sum_{i=1}^mV_i$, where $V_i=(\bb_i^T\bw)sgn(K_i-\bb_i^T\bw)$. By Chebyshev's inequality, $P(V<E(V)+h\times \SD(V))>1-h^{-2}$. Therefore, to prove the result in Theorem 2, we want to derive the upper bounds for $E(V)$ and $\SD(V)$ and show that $\forall h>0$, $\exists B$ and $M$ s.t. $\forall m>M$, $P(V<0)>1-h^{-2}$.

We will first present a result from Polya (1945), which will be very useful for our proof. For $x>0$,
\begin{equation}\label{a70}
\Phi(x)=\frac{1}{2}\Big[1+\sqrt{1-\exp(-\frac{2}{\pi}x^2)}\Big](1+\delta(x)) \ \ \ \text{with} \ \ \sup_{x>0}|\delta(x)|<0.004.
\end{equation}
The variance of $V$ is shown as follows:
\begin{equation*}
\Var(V)=m^{-2}\sum_{i=1}^m\Var(V_i)+m^{-2}\sum_{i\neq j}\Cov(V_i,V_j).
\end{equation*}
Write $\bw=s\bu$ with $\|\bu\|_2=1$ where $s=(Bk/m)^{1/2}$.  By (C2), (C3) and (C4) in Theorem 2, for sufficiently large $m$,
\begin{eqnarray}\label{d1}
\sum_{i=1}^m\Var(V_i)&=&\sum_{i=1}^mI(|\bb_i^T\bu|\leq d)\Var(V_i)+\sum_{i=1}^mI(|\bb_i^T\bu|>d)\Var(V_i)\nonumber \\
                 &=&\Big[\sum_{i=1}^mI(|\bb_i^T\bu|>d)\Var(V_i)\Big](1+o(1)),
\end{eqnarray}
and
\vspace{-0.3cm}
\begin{eqnarray}\label{d2}
\sum_{i\neq j}\Cov(V_i,V_j)&=&\sum_{i\neq j}I(|\bb_i^T\bu|\leq d)I(|\bb_j^T\bu|\leq d)\Cov(V_i,V_j)\nonumber \\
     &&+2\sum_{i\neq j}I(|\bb_i^T\bu|\leq d)I(|\bb_j^T\bu|>d)\Cov(V_i,V_j)\nonumber\\
     &&+\sum_{i\neq j}I(|\bb_i^T\bu|>d)I(|\bb_j^T\bu|>d)\Cov(V_i,V_j)\nonumber\\
     &=&\Big[\sum_{i\neq j}I(|\bb_i^T\bu|>d)I(|\bb_j^T\bu|>d)\Cov(V_i,V_j)\Big](1+o(1)).
\end{eqnarray}
We will prove (\ref{d1}) and (\ref{d2}) in detail at the end of proof for Theorem 2.

For each pair of $V_i$ and $V_j$, it is easy to show that
\begin{equation*}
\Cov(V_i,V_j)=4(\bb_i^T\bw)(\bb_j^T\bw)\Big[P(K_i<\bb_i^T\bw,K_j<\bb_j^T\bw)-\Phi(a_i\bb_i^T\bw)\Phi(a_i\bb_j^T\bw)\Big].
\end{equation*}
The above formula includes the $\Var(V_i)$ as a specific case.

By Polya's approximation (\ref{a70}),
\begin{equation}\label{d3}
\Var(V_i)=(\bb_i^T\bw)^2\exp\Big\{-\frac{2}{\pi}(a_i\bb_i^T\bw)^2\Big\}(1+\delta_j) \ \ \text{with} \ |\delta_j|<0.004.
\end{equation}
Hence
\begin{eqnarray*}
\sum_{i=1}^mI(|\bb_i^T\bu|>d)\Var(V_i)&\leq&\sum_{i=1}^ms^2\exp\Big\{-\frac{2}{\pi}(a_ids)^2\Big\}(1+\delta_j)\\
                    &\leq&2ms^2\exp\Big\{-\frac{2}{\pi}(a_{\min}ds)^2\Big\}.
\end{eqnarray*}
To compute $\Cov(V_i,V_j)$, we have
\begin{eqnarray*}
&&P(K_i<\bb_i^T\bw,K_j<\bb_j^T\bw)\\
&=&\int_{-\infty}^{\infty}\Phi\Big(\frac{(|\rho_{ij}^k|)^{1/2}z+a_i\bb_i^T\bw}{(1-|\rho_{ij}^k|)^{1/2}}\Big)\Phi\Big(\frac{\delta_{ij}^k(|\rho_{ij}^k|)^{1/2}z+a_j\bb_j^T\bw}{(1-|\rho_{ij}^k|)^{1/2}}\Big)\phi(z)dz\\
&=&\Phi(a_i\bb_i^T\bw)\Phi(a_j\bb_j^T\bw)+\phi(a_i\bb_i^T\bw)\phi(a_j\bb_j^T\bw)a_ia_jcov_{ij}^k(1+o(1)),
\end{eqnarray*}
where $\delta_{ij}^k=1$ if $\rho_{ij}^k\geq0$ and $-1$ otherwise. Therefore,
\begin{equation}\label{d4}
\Cov(V_i,V_j)= 4(\bb_i^T\bw)(\bb_j^T\bw)\phi(a_i\bb_i^T\bw)\phi(a_j\bb_j^T\bw)a_ia_jcov_{ij}^k(1+o(1)),
\end{equation}
and
\begin{eqnarray*}
&&|\sum_{i\neq j}I(|\bb_i^T\bu|>d)I(|\bb_j^T\bu|>d)\Cov(V_i,V_j)|\\
&<&\sum_{i\neq j}s^2\exp\Big\{-(a_{\min}ds)^2\Big\}a_{\max}^2|cov_{ij}^k|(1+o(1)).
\end{eqnarray*}
Consequently, we have
\begin{equation*}
\Var(V)<\frac{2}{m}s^2\exp\Big\{-\frac{2}{\pi}(a_{\min}ds)^2\Big\}a_{\max}^2\Big[\frac{1}{m}\sum_i\sum_j|cov_{ij}^k|\Big].\\
\end{equation*}
We apply (C1) in Theorem 2 and the Cauchy-Schwartz inequality to get $\frac{1}{m}\sum_i\sum_jcov_{ij}^k\leq(\sum_{i=k+1}^p\lambda_i^2)^{1/2}$ $\leq\eta^{1/2}$, and conclude that the standard deviation of $V$ is bounded by
\begin{equation*}
\sqrt{2}sm^{-1/2}\exp\Big\{-\frac{1}{\pi}(a_{\min}ds)^2\Big\}a_{\max}(\eta)^{1/4}.
\end{equation*}
In the derivations above, we used the fact that $\bb_i^T\bu\leq\|\bb_i\|_2<1$, and the covariance matrix for $K_i$ in (21) of the paper is a submatrix for covariance matrix of $K_i$ in (10).

Next we will show that $E(V)$ is bounded from above by a negative constant. Using $x(\Phi(x)-\frac{1}{2})\geq0$, we have
\begin{eqnarray*}
-E(V)&=&\frac{2}{m}\sum_{i=1}^m\bb_i^T\bw\Big[\Phi(a_i\bb_i^T\bw)-\frac{1}{2}\Big]\\
    &\geq&\frac{2ds}{m}\sum_{i=1}^mI(|\bb_i^T\bu|>d)\Big[\Phi(a_ids)-\frac{1}{2}\Big]\\
    &=&\frac{2ds}{m}\sum_{i=1}^m\Big[\Phi(a_ids)-\frac{1}{2}\Big]-\frac{2ds}{m}\sum_{i=1}^mI(|\bb_i^T\bu|\leq d)\Big[\Phi(a_ids)-\frac{1}{2}\Big].
\end{eqnarray*}
By (C2) in Theorem 2, $\frac{1}{m}\sum_{i=1}^mI(|\bb_i^T\bu|\leq d)\rightarrow0$, so for sufficiently large $m$, we have
\begin{equation*}
-E(V)\geq \frac{ds}{m}\sum_{i=1}^m\Big[\Phi(a_ids)-\frac{1}{2}\Big].
\end{equation*}
An application of (\ref{a70}) to the right hand side of the last line leads to
\begin{eqnarray*}
-E(V)\geq \frac{ds}{m}\sum_{i=1}^m\frac{1}{2}\sqrt{1-\exp\big\{-\frac{2}{\pi}(a_{\min}ds)^2\big\}}.
\end{eqnarray*}
Note that
\begin{equation*}
1-\exp(-\frac{2}{\pi}x^2)=\frac{2}{\pi}x^2\sum_{l=0}^{\infty}\frac{1}{(l+1)!}(-\frac{2}{\pi}x^2)^l>\frac{2}{\pi}x^2\sum_{l=0}^{\infty}\frac{1}{l!}(-\frac{1}{\pi}x^2)^l=\frac{2}{\pi}x^2\exp(-\frac{1}{\pi}x^2),
\end{equation*}
so we have
\begin{equation*}
-E(V)\geq \frac{d^2}{2}s^2\sqrt{\frac{2}{\pi}}a_{\min}\exp\Big\{-\frac{1}{2\pi}(a_{\min}ds)^2\Big\}.
\end{equation*}
\indent To show that $\forall h>0$, $\exists B$ and $M$ s.t. $\forall m>M$, $P(V<0)>1-h^{-2}$, by Chebyshev's inequality and the upper bounds derived above, it is sufficient to show that
\begin{equation*}
\frac{d^2}{2}s^2\sqrt{\frac{1}{\pi}}a_{\min}\exp\Big\{-\frac{1}{2\pi}(a_{\min}ds)^2\Big\}>hsm^{-1/2}\exp\Big\{-\frac{1}{\pi}(a_{\min}ds)^2\Big\}a_{\max}\eta^{1/4}.
\end{equation*}
Recall $s=(Bk/m)^{1/2}$, after some algebra, this is equivalent to show
\begin{equation*}
d^2(Bk)^{1/2}(\pi)^{-1/2}\exp\Big\{\frac{1}{2\pi}(a_{\min}ds)^2\Big\}>2h\eta^{1/4}\frac{a_{\max}}{a_{\min}}.
\end{equation*}
By (C3), then for all $h>0$, when $B$ satisfies $d^2(Bk)^{1/2}(\pi)^{-1/2}>2h\eta^{1/4}S$, we have $P(V<0)>1-h^{-2}$. Note that $k=O(m^{\kappa})$ and $\eta=O(m^{2\kappa})$, so $k^{-1/2}\eta^{1/4}=O(1)$. To complete the proof of Theorem 2, we only need to show that (\ref{d1}) and (\ref{d2}) are correct.

To prove (\ref{d1}), by (\ref{d3}) we have
\begin{equation*}
\sum_{i=1}^mI(|\bb_i^T\bu|\leq d)\Var(V_i)\leq\sum_{i=1}^mI(|\bb_i^T\bu|\leq d)s^2d^2,
\end{equation*}
and
\begin{equation*}
\sum_{i=1}^mI(|\bb_i^T\bu|   > d)\Var(V_i)\geq\sum_{i=1}^mI(|\bb_i^T\bu|   > d)s^2d^2\exp\Big\{-\frac{2}{\pi}a_{\max}^2s^2\Big\}.
\end{equation*}
Recall $s=(Bk/m)^{1/2}$, then by (C3) and (C4), $\exp\Big\{\frac{2}{\pi}a_{\max}^2s^2\Big\}=O(1)$. Therefore, by (C2) we have
\begin{equation*}
\frac{\sum_{i=1}^mI(|\bb_i^T\bu|\leq d)\Var(V_i)}{\sum_{i=1}^mI(|\bb_i^T\bu|   > d)\Var(V_i)}\rightarrow0 \ \ \text{as} \ \ m\rightarrow\infty,
\end{equation*}
so (\ref{d1}) is correct. With the same argument and by (\ref{d4}), we can show that (\ref{d2}) is also correct. The proof of Theorem 2 is now complete.

\noindent\textbf{Proof of Theorem 3:} Letting
\begin{eqnarray*}
\Delta_1&=&\sum_{i=1}^p\Big[\Phi(a_i(z_{t/2}+\bb_i^T\hw))-\Phi(a_i(z_{t/2}+\bb_i^T\bw))\Big]\quad\quad\text{and}\\
\Delta_2&=&\sum_{i=1}^p\Big[\Phi(a_i(z_{t/2}-\bb_i^T\hw))-\Phi(a_i(z_{t/2}-\bb_i^T\bw))\Big],
\end{eqnarray*}
we have
\begin{equation*}
\widehat{\FDP}(t)-\FDP_A(t)=(\Delta_1+\Delta_2)/R(t).
\end{equation*}
Consider $\Delta_1=\sum_{i=1}^p\Delta_{1i}$. By the mean value theorem, there exists $\xi_i$ in the interval of $(\bb_i^T\hw,\bb_i^T\bw)$, such that
$\Delta_{1i}=\phi(a_i(z_{t/2}+\xi_i))a_i\bb_i^T(\hw-\bw)$ where $\phi(\cdot)$ is the standard normal density function.

Next we will show that $\phi(a_i(z_{t/2}+\xi_i))a_i$ is bounded by a constant. Without loss of generality, we discuss about the case in (C6) when $z_{t/2}+\bb_i^T\bw<-\tau$. By Theorem 2, we can choose sufficiently large $m$ such that $z_{t/2}+\xi_i<-\tau/2$. For the function $g(a)=\exp(-a^2x^2/8)a$, $g(a)$ is maximized when $a=2/x$. Therefore,
\begin{equation*}
\sqrt{2\pi}\phi(a_i(z_{t/2}+\xi_i))a_i<a_i\exp(-a_i^2\tau^2/8)\leq2\exp(-1/2)/\tau.
\end{equation*}
For $z_{t/2}+\bb_i^T\bw>\tau$ we have the same result. In both cases, we can use a constant $D$ such that $\phi(a_i(z_{t/2}+\xi_i))a_i\leq D$.

By the Cauchy-Schwartz inequality, we have $\sum_{i=1}^p|b_{ih}|\leq(p\sum_{i=1}^pb_{ih}^2)^{1/2}=(p\lambda_h)^{1/2}$. Therefore, by the Cauchy-Schwartz inequality and the fact that $\sum_{h=1}^k\lambda_h<p$, we have
\begin{eqnarray*}
|\Delta_{1}|&\leq&D\sum_{i=1}^p\Big[\sum_{h=1}^k|b_{ih}||\widehat{w}_h-w_h|\Big]\\
       &\leq&D\sum_{h=1}^k(p\lambda_h)^{1/2}|\widehat{w}_h-w_h|\\
       &\leq&D\sqrt{p}\Big(\sum_{h=1}^k\lambda_h\sum_{h=1}^k(\widehat{w}_h-w_h)^2\Big)^{1/2}\\
       &<&Dp\|\hw-\bw\|_2.
\end{eqnarray*}
By Theorem 1, $\|\hw-\bw\|_2=O_p(\sqrt{\frac{k}{m}})$. By (C5) in Theorem 3, $R(t)/p>H$ for $H>0$ when $p\rightarrow\infty$. Therefore, $|\Delta_{1}/R(t)|=O_p(\sqrt{\frac{k}{m}})$. For $\Delta_{2}$, the result is the same. The proof of Theorem 3 is now complete.

\noindent\textbf{Proof of Theorem 4:} Note that $\|\widehat{\bW}_{\LS}-\widehat{\bW}_{\LS}^*\|_2=\|(\bX^T\bX)^{-1}\bX^T\bmu\|_2$. By the definition of $\bX$, we have $\bX^T\bX=\Lambda$, where $\Lambda=\diag(\lambda_1,\cdots,\lambda_k)$. Therefore, by the Cauchy-Schwartz inequality,
\begin{equation*}
\|\widehat{\bW}_{\LS}-\widehat{\bW}_{\LS}^*\|_2=\Big[\sum_{i=1}^k\big(\frac{\sqrt{\lambda_i}\bgamma_i^T\bmu}{\lambda_i}\big)^2\Big]^{1/2}\leq\|\bmu\|_2\Big(\sum_{i=1}^k\frac{1}{\lambda_i}\Big)^{1/2}
\end{equation*}
The proof is complete.

%\bibliographystyle{asa}
%\bibliography{mainbibyy}
\bibliographystyle{ims}
\bibliography{amsis-ref}
\end{document}